\documentclass[]{aastex631}
\usepackage{amsmath}
\usepackage{siunitx}

\newcommand{\BV}{Brunt-Väisälä }
\newcommand{\BVfreq}{Brunt-Väisälä frequency }

\received{March 8, 2023}
\accepted{August 7, 2023}

\submitjournal{PSJ}

\begin{document}

\title{Effects of the librationally induced flow in Mercury's fluid core with an outer stably stratified layer}

\correspondingauthor{Fleur Seuren}
\email{fleur.seuren@ksb-orb.be}

\author[0000-0002-7128-8764]{Fleur Seuren}
\affiliation{Royal Observatory of Belgium, Ringlaan 3, 1180 Brussels, Belgium}
\affiliation{Institute of Astronomy, KU Leuven, Celestijnenlaan 200D, 3001 Leuven, Belgium}

\author[0000-0002-7679-3962]{Santiago A. Triana}
\affiliation{Royal Observatory of Belgium, Ringlaan 3, 1180 Brussels, Belgium}

\author[0000-0003-3151-6969]{Jérémy Rekier}
\affiliation{Royal Observatory of Belgium, Ringlaan 3, 1180 Brussels, Belgium}

\author[0000-0001-5747-669X]{Ankit Barik}
\affiliation{Johns Hopkins University, 3400 N. Charles Street, Baltimore, MD 21210, USA}

\author[0000-0002-9820-8584]{Tim Van Hoolst}
\affiliation{Royal Observatory of Belgium, Ringlaan 3, 1180 Brussels, Belgium}
\affiliation{Institute of Astronomy, KU Leuven, Celestijnenlaan 200D, 3001 Leuven, Belgium}




\begin{abstract}
Observational constraints on Mercury's thermal evolution and magnetic field indicate that the top part of the fluid core is stably stratified. Here we compute how a stable layer affects the core flow in response to Mercury's main 88-day longitudinal libration, assuming various degrees of stratification, and study whether the core flow can modify the libration amplitude through viscous and electromagnetic torques acting on the core-mantle boundary (CMB). We show that the core flow strongly depends on the strength of the stratification near the CMB but that the influence of core motions on libration is negligible with or without a stably stratified layer. A stably stratified layer at the top of the core can however prevent resonant behaviour with gravito-inertial modes by impeding radial motions and promote a strong horizontal flow near the CMB. The librationally driven flow is likely turbulent and might produce a non-axisymmetric induced magnetic field with a strength of the order of $1\%$ of Mercury's dipolar field.
\end{abstract}




\section{Introduction} \label{sec:intro}
Mercury experiences small periodic variations in its rotation rate called librations \citep[][]{margot2007large}. It is generally considered that Mercury's fluid outer core rotates with constant speed and does not show similar rotational variations \citep[e.g. in][]{peale2002procedure, vanhoolst2012effect}. Part of the angular momentum of the solid mantle can however be transferred to the liquid outer core via various core-mantle coupling mechanisms such as the viscous drag at the core-mantle boundary (CMB) and the electromagnetic torque produced by the Lorentz force at the base of the conducting mantle \citep[see e.g.][]{dehant2015earthrotation}, generating fluid motions in the fluid interior \citep[see][for a review]{lebars2015flows}. In the present study we focus on the core flow induced by Mercury's forced 88-day longitudinal libration, which is within the frequency range of inertial modes \citep[see e.g.][]{tilgner1999driven}. 

It is important to understand the exact nature of the librationally induced flows as internal fluid motions can influence the magnetic field and the rotational dynamics (including libration itself), both of which are used to constrain properties of Mercury's interior, see e.g. \citet[][]{wardinski2021internal} for the former and e.g. \citet[][]{margot2007large,vanhoolst2012effect} for the latter. Here we consider recent developments in thermal evolution models \citep[e.g.][]{knibbe2018thermal, knibbe2021modelling} as well as in numerical dynamo models \citep[e.g.][]{christensen2006deep, tian2015magnetic, takahashi2019mercury} that indicate the presence of a thick stable layer at the top of Mercury's core, which potentially influences the flow near the boundary. This is something that has not been considered in previous experimental \citep[e.g.][]{aldridge1969axisymmetric, noir2009experimental} or numerical \citep[e.g.][]{calkins2010axisymmetric,lin2020libration} studies into Mercury's fluid core response to its longitudinal libration, nor in theoretical studies that estimate that the influence of viscous \citep[e.g.][]{peale2002procedure} and electromagnetic coupling \citep[e.g.][]{peale2002procedure, dumberry2011free} on the libration amplitude is negligible, based on the rotation and estimated diffusion time scales. 

Our first motivation to examine the librationally induced flow, influenced by an outer stably stratified layer, is to investigate the latter claim by calculating the effect of the outer core flow on the libration amplitude. Studies that estimate the inner and outer core sizes or the core and mantle densities from libration observations typically consider the outer core to be in hydrostatic equilibrium and assume that the flow in the outer core does not affect the libration amplitude \citep[][]{vanhoolst2012effect, hauck2013curious, dumberry2013role, rivoldini2013interior, steinbruegge2021challenges, knibbe2021modelling, lark2022sulfides}. Here we use for the first time an explicit calculation of both the outer core motions and the resulting torques on Mercury's mantle to assess whether the torques and the libration amplitude can be significantly affected through for example a resonant excitation of an eigenmode. Second, we investigate in detail the influence of a stably stratified layer on the librationally induced flow and on the associated induced magnetic field. In the Earth's core, where the existence of a similar but much thinner and weaker stratified layer has been hypothesised, numerical studies have shown that stratification can indeed significantly change the fluid flow near the outer boundary by limiting the penetration of quasi-toroidal (columnar) modes into that layer \cite[][]{takehiro2001penetration, nakagawa2011effect, vidal2015quasi} and hosting Magneto-Archimedes-Coriolis (MAC) waves \citep[][]{braginsky1999dynamics} that could account for fluctuations in the magnetic field \citep[][]{buffett2014geomagnetic} and length-of-day observations \citep[][]{buffett2016evidence}. 

In this study we take a first step to investigate the possible effects of an outer stably stratified layer on the librationally induced flow in Mercury's outer core. We follow \citet[][]{rekier2019internal} and numerically compute the linear flow response to a harmonic oscillation of the spherical outer boundary, using the linear model presented in Section \ref{sec:model} and the numerical method presented in Section \ref{sec:method}. The choice for a linear model over a nonlinear model allows us to reach parameter regimes much closer to the expected values in Mercury's outer core than in previous studies \citep[e.g.][]{noir2009experimental}, but neglects at the same time any turbulent effects which we will reflect on in later sections. In Section \ref{sec:theory} we consider how the core angular momentum can change from a theoretical point of view. In Section \ref{sec:torque} we use our model to compute the actual viscous and electromagnetic torques from the core on the mantle to see the influence of the core flow on the mantle rotation and in Section \ref{sec:results} we highlight the different ways an outer stably stratified layer can influence the core motions and the induced magnetic field.  In Section \ref{sec:conclusions} we present our conclusions. 

\section{A simple model for the librationally induced flow in Mercury's core} \label{sec:model}
\subsection{Principal equations} \label{sec:equations}
We consider the motion of a homogeneous, viscous, and electrically conductive fluid in a rotating spherical shell with inner radius $r_\mathrm{ICB}$, corresponding to Mercury's inner core boundary, and outer radius $r_\mathrm{CMB}$, corresponding to Mercury's core-mantle boundary. We adopt the Boussinesq approximation, and consider variations in density that result from thermal effects:
\begin{equation}
\rho = \bar{\rho}(1 - \alpha (T - \bar{T}))~. \label{eq:density}
\end{equation}
Here $\alpha$ is the coefficient of thermal expansion and $\bar{\rho}$ and $\bar{T}$ are constant spatial averages of the density $\rho$ and temperature $T$. In the Boussinesq approach, variations in density are neglected in the equations for conservation of mass and momentum, except in the buoyancy term where they are multiplied with the gravity acceleration term $\mathbf{g}$. Ignoring any gravity perturbations and assuming a homogeneous fluid in the outer core with the same density as the inner core, $\mathbf{g}$ only depends on the radial coordinate $r$:
\begin{equation}
\mathbf{g} = -\frac{g_0r}{r_\mathrm{CMB}}\mathbf{\hat{r}}~,
\end{equation}
where $g_0$ is the acceleration of gravity at the core-mantle boundary and $\hat{\mathbf{r}}$ denotes the unit vector in the radial direction. In the absence of any fluid flow the temperature and magnetic field are held constant, as they are generated by processes that occur on much longer timescales than the libration that we are interested in. We write the background temperature, $\bar{T} + T_0(r)$, as a radial function whose gradient interpolates between 0 and some positive value, so that we treat any convective regions as well-mixed with zero temperature gradient and consider only the action of the buoyancy force in stably stratified regions (i.e. with a positive temperature gradient). We further take the background magnetic field $\mathbf{B}_0$ to be dipolar:
 \begin{equation}
 \mathbf{B}_0 = \mathbf{\nabla \times \nabla \times} \left[\frac{B_0}{2r^2}\mathrm{Y}_1^0(\theta, \phi) \mathbf{r} \right]~,\label{eq:dipole}
 \end{equation}
with constant $B_0$ and $\mathrm{Y}_1^0$, the Schmidt semi-normalised spherical harmonic (i.e. the standard spherical harmonic with normalisation factor $\sqrt{(\ell-m)!/(\ell+m)!}$) with degree $\ell = 1$ and order $m = 0$. The real configuration of Mercury's magnetic field is of course infinitely more complex \citep[see e.g.][]{anderson2012low} and is for example much stronger in the middle of the fluid core \citep[see e.g. the dynamo model in][]{christensen2006deep}. We nevertheless opt here for a simple dipolar structure of the background field since the background magnetic field has only a limited influence on our results (see also Appendix \ref{app:sensitivity}).

As is standard practice \citep[see e.g.][for reviews]{finlay2008waves, triana2021core} the linearized outer core dynamics in a reference frame rotating with Mercury's mean angular velocity $\mathbf{\Omega} = \Omega\hat{\mathbf{z}}$ can be described by small perturbations of the flow velocity $\mathbf{u}$, magnetic field $\mathbf{b}$, and temperature $\Theta$ around a steady background state $\mathbf{U}_0 = 0, \mathbf{B}_0, \bar{T} + T_0(r)$, resulting in the usual Boussinesq MHD equations \citep[see e.g.][]{davidson2001magnetohydrodynamics}:
\begin{equation}
\partial_t \mathbf{u} + 2\mathbf{\Omega \times u}= -\mathbf{\nabla} \left(\frac{P}{\bar{\rho}}-\frac{1}{2}|\mathbf{\Omega}\times\mathbf{r}|^2+\Phi\right) + \alpha \Theta \mathbf{g} + \nu \mathbf{\nabla}^2 \mathbf{u} + {(\mu_0\bar{\rho})}^{-1}\left[(\mathbf{\nabla \times b}) \mathbf{ \times } \mathbf{B}_0 \right] ~, \label{eq:momentum}
\end{equation}
\begin{equation}
\partial_t \mathbf{b} = \mathbf{\nabla \times} (\mathbf{u \times B}_0) + \eta \mathbf{\nabla}^2 \mathbf{b}~, \label{eq:induction}
\end{equation}
\begin{equation}
\partial_t \Theta = u_r\partial_r T_0 + \kappa \mathbf{\nabla}^2\Theta~, \label{eq:temperature}
\end{equation}
\begin{equation}
\mathbf{\nabla \cdot u} = \mathbf{0}~. \label{eq:incompressible}
\end{equation}
where $P$ is the pressure, $\mu_0$ the vacuum permeability, $\nu$, $\eta$, and $\kappa$, respectively, denote the kinematic viscosity, magnetic diffusivity and thermal diffusivity and $\Phi$ denotes the gravitational potential. Note that a static background flow ($\mathbf{U}_0 = 0$) is an appropriate approximation in our linearized approach as illustrated by the smallness of the Rossby number associated with large-scale convective flows \citep[typically of the order of 0.001,][]{olson2006dipole} and we can consider the small librationally induced flows over an 88-day period separate from the small flows related to the dynamo generation which evolve over much longer time scales. Since we do not consider any nonlinear interactions in Equations \eqref{eq:momentum}-\eqref{eq:incompressible}, we will not be able observe any non-linear instabilities or turbulent effects. On the other hand the linear model allows us to probe much deeper into the parameter regime than previous studies \citep[e.g.][]{noir2009experimental, calkins2010axisymmetric}, revealing previously unseen characteristics of the librationally induced flow.

In the mean rotating frame, the only motion of the inner and outer boundaries are due to deviations from Mercury's mean rotation, that we assume here to be Mercury's longitudinal librations as first observed by \citet[][]{margot2007large}. Libration causes the mantle and thus the outer boundary of the core to oscillate with a velocity $\mathbf{v}$ in the azimuthal direction. It also causes the inner core to oscillate, but as the libration of the inner core is expected to be about a factor 20 smaller than the libration of the mantle \citep[][]{vanhoolst2012effect} we neglect this and assume that the solid inner core spins with the mean rotation. 

At both boundaries the flow velocity must satisfy the no-slip conditions:
\begin{equation}
\mathbf{u}|_{r = r_\mathrm{ICB}} = \mathbf{0}~,  \hspace{4cm} \mathbf{u}|_{r = r_\mathrm{CMB}} = \mathbf{v}~ \label{eq:no_slip}
\end{equation}
and we further impose that there is no additional heat flux into either the mantle or the solid inner core resulting from temperature perturbations $\Theta$.
\begin{equation}
\left.\frac{d\Theta}{dr}\right|_{r = r_\mathrm{ICB}} =  \left.\frac{d\Theta}{dr}\right|_{r = r_\mathrm{CMB}} = 0~. \label{eq:constant_heatflux}
\end{equation}
In Equations \eqref{eq:no_slip} and \eqref{eq:constant_heatflux}), we treat the boundaries as spherical and located at radii $r_\mathrm{ICB}$ and $r_\mathrm{CMB}$. In reality, these boundaries are flattened at the poles and elongated at the equator forming non-axisymmetric shapes whose pressure on the liquid core produces a flow motion during libration (see \ref{sec:librating_boundaries} below).

To allow for electromagnetic coupling between the core and the mantle, we consider a thin electrically conducting layer at the base of the mantle with width $\delta$ and magnetic diffusivity $\eta_W$, and use a thin-wall boundary condition as the outer magnetic boundary condition on the interface between the conducting layer and the outer core (see Appendix \ref{app:poltor_bound}). The thin-wall condition is governed by two parameters $c$ and $c'$, where $c = \mu_W \delta / (\mu_F r_\mathrm{CMB})$ controls the ratio between the magnetic permeability in the thin layer $\mu_W$ and the permeability in the core $\mu_F$, and $c' = \sigma_W \delta / (\sigma_F r_\mathrm{CMB})$ is a conductance ratio between mantle and core, defined by the ratio of the electrical conductivity of the base of the mantle $\sigma_W$ and that of the core fluid $\sigma_F$. The condition is only valid in the thin-wall approximation (TWA), when $\delta$ is much smaller than the generalised skin depth of the mantle $\mathcal{\delta}_\eta$ \citep[][]{roberts2010numerical} but in Appendix \ref{app:poltor_bound} we show that this requirement is met for all realistic models of Mercury and its libration.

Since the nature of the magnetic boundary condition on the inner surface turns out to have very little influence on our results, see Figure \ref{fig:testB} in Appendix \ref{app:sensitivity}, we impose on this boundary the numerically advantageous insulating boundary condition, in which the magnetic field matches a potential field in the solid inner core.

\subsection{Librating boundaries} \label{sec:librating_boundaries}
Similar to \citet[][]{rekier2019internal} we represent the librational motion $\mathbf{v}$ of the spherical outer boundary as the superposition of three different oscillating motions. The first oscillation, an angular displacement in the longitudinal direction $\phi \rightarrow \phi + \zeta(t)$, is directly caused by the $z$ component of the solar gravitational torque. For a harmonic oscillation $\zeta(t) = \epsilon \cos\omega t$ the tangential velocity $v_\phi = s\dot{\phi}$ of the oscillating boundary, with cylindrical radius $s = r_\mathrm{CMB}\sin\theta$ and co-latitude $\theta$, becomes:
\begin{equation}
v_\phi = s\dot{\zeta} = -s\epsilon\omega \sin{\omega t}~ =  i r_\mathrm{CMB} \sin\theta \,\omega \frac{\epsilon}{2}\left(e^{i\omega t} - e^{-i\omega t}\right)~, \label{eq:vel_phi}
\end{equation}
where $\epsilon$ is the libration amplitude, about $\SI{39}{\arcsecond}$ or $1.89 \times 10^{-4}~\SI{}{\radian}$, and $\omega$ is the libration frequency, approximately $0.67\Omega$ \citep[][]{margot2012mercury, stark2015first, genova2019geodetic, bertone2021deriving}. We will refer to this oscillation as the $m=0$ component of libration. 

The second component of libration follows from the fact that Mercury actually has a slightly triaxial shape, defined by the equatorial axes with lengths $a > b$ and the polar axis with length $c$. The azimuthal librating motion will slightly change the radial coordinate of each point on the core-mantle boundary. We can imitate this in our spherical model by writing the radial displacement of the triaxial boundary as a radial in- and outflow of the spherical boundary. To first order, the radial coordinate of the moving triaxial boundary can be described via a spherical harmonic expansion as:
\begin{equation}
\frac{r(\theta, \phi)}{r_\mathrm{CMB}} = 1 + \alpha_0 \mathrm{Y}^0_2(\theta, \phi) + \alpha_2 \left[\mathrm{Y}^2_2(\theta, \phi) + \mathrm{Y}_2^{-2}(\theta, \phi) \right], \label{eq:shape}
\end{equation}
so that the radial velocity $v_r$ of the librating boundary can be expressed as:
\begin{equation}
v_r(t) = \dot{r}(t)=2i r_\mathrm{CMB}\alpha_2\beta_2\, \dot{\zeta}\left(  \mathrm{e}^{2i(\phi+\zeta)} -  \mathrm{e}^{-2i(\phi+\zeta)} \right)~, 
\end{equation}
where $\mathrm{Y}_\ell^m(\theta, \phi) = \beta_\ell e^{im\phi}$ are again the Schmidt semi-normalised spherical harmonics, $\beta_\ell$ equals the Legendre polynomial $P_\ell^m(\theta)$ multiplied by the normalisation factor $\sqrt{(\ell-m)!/(\ell+m)!}$, and $\alpha_0$, $\alpha_2$ are real numbers describing the polar and equatorial flattening:
\begin{equation}
\alpha_0 = \frac{(a + b) - 2c}{(a + b)}~,  \hspace{4cm} \alpha_2 = \frac{a - b}{a}~. \label{eq:flattening}
\end{equation}
To first order in the libration amplitude $\epsilon$ this can be approximated by:
\begin{equation}
v_r = -r_\mathrm{CMB} \alpha_2 \beta_2\, \omega \epsilon  \left[ \left(  \mathrm{e}^{i (2\phi + \omega t)} +  \mathrm{e}^{-2i(\phi + \omega t)}  \right) -  \left(  \mathrm{e}^{i( 2\phi-\omega t)} +  \mathrm{e}^{-i( 2 \phi-\omega t}     \right)\right]~, \label{eq:vel_rad}
\end{equation}
which only depends on the azimuthal orders $m=\pm 2$. The radial velocity can then be written as the sum of two components $f(\omega)$ and $f(-\omega)$ with the function $f(\omega) = f(\theta, \phi, \omega,t)$ given as:
\begin{equation}
    f(\theta,\phi,\omega,t) \equiv -r_\mathrm{CMB}\alpha_2 \beta_2\, \omega \epsilon \left(  \mathrm{e}^{i (2\phi + \omega t)} +  \mathrm{e}^{-i(2\phi + \omega t)}  \right) = -r_\mathrm{CMB}\alpha_2\omega\epsilon\,\mathrm{Y}_2^2(\theta,\phi)\mathrm{e}^{i\omega t} + \mathrm{c.c.}~,
\end{equation}
where the complex conjugate is indicated by $\mathrm{c.c.}$. We call $v_r = f(\omega)$ the $m=2$ component and $v_r = f(-\omega)$ the $m=-2$ component of libration. 

In summary, we describe the libration forcing in the sphere as the sum of three harmonic components $m=0$ (axial forcing), and $m=\pm 2$ (radial forcings). The spherical symmetry of the computational domain preserved in our model implies that groups of variables with different $m$-numbers are decoupled from each other, allowing us to treat each forcing component independently. An advantage of this simplification is that it allows us to reach parameters values that are closer to the expected values in Mercury. The error it introduces is of the second order in the small flattening parameter $\alpha_2$. We can see this by considering Equations~\eqref{eq:no_slip}, replacing $r_\mathrm{CMB}$ by $r(\theta,\phi)$ of Equation \eqref{eq:shape}, and expanding the resulting expression in series of $\alpha_2$. The result then follows once we note that $\mathbf{v}$, given by Equation \eqref{eq:vel_rad}, is itself of the first order in $\alpha_2$.

\section{Numerical implementation} \label{sec:method}
\subsection{Non-dimensional equations} \label{sec:nd_equations}
We determine the outer core flow in response to the mantle libration by solving a non-dimensional version of the Boussinesq MHD equations \eqref{eq:momentum}-\eqref{eq:incompressible}. We set the outer core radius $r_\mathrm{CMB}$ as the unit of length $R$ and the rotation time scale $\tau_\Omega = \Omega^{-1}$ as the unit of time. By using $\bar{\rho} R^2 \Omega^2$ for the pressure scale,  $R \Omega^2 / \alpha g_0$ for the temperature scale, and $B_0$ for the magnetic field scale, Equations \eqref{eq:momentum}-\eqref{eq:incompressible} can be rewritten to:
\begin{equation}
\partial_t \mathbf{u} + 2 \hat{\mathbf{z}} \times \mathbf{u} = - \mathbf{\nabla} p + \Theta\mathbf{r} + \mathrm{Ek} \mathbf{\nabla}^2 \mathbf{u} + \mathrm{Le}^2\left[(\mathbf{\nabla} \times \mathbf{b} ) \times \mathbf{B}_0\right]~, \label{eq:momentum_nd} 
\end{equation}
\begin{equation}
\partial_t \mathbf{b} = \mathbf{\nabla} \times \left(\mathbf{u} \times \mathbf{B}_0 \right) + \mathrm{Em}\mathbf{\nabla}^2 \mathbf{b}~, \label{eq:induction_nd} 
\end{equation}
\begin{equation}
\partial_t \Theta = - N^2(r) u_r + \left(\mathrm{Ek}/\mathrm{Pr}\right) \mathbf{\nabla}^2\Theta~, \label{eq:temperature_nd}
\end{equation}
\begin{equation}
\mathbf{\nabla}\cdot\mathbf{u} = 0~. \label{eq:incompressible_nd}
\end{equation}
where we have introduced the (non-dimensional) reduced pressure, $p=P/\bar{\rho}-|\hat{\mathbf{z}}\times\mathbf{r}|^2/2 - \Phi$.

The non-dimensional system of equations is governed by a radial function of the squared dimensionless \BVfreq $N^2(r)$ and four other dimensionless parameters. The Ekman number $\mathrm{Ek} = \nu/\Omega R^2$ sets the ratio between the rotation time scale and the viscous diffusion time scale $\tau_\nu = R^2\nu^{-1}$, the Lehnert number $\mathrm{Le} = B_0/\sqrt{\bar{\rho}\mu_0}\Omega R$ represents the ratio between the rotation time scale and the Alfvén wave time scale $\tau_A = R\sqrt{\bar{\rho}\mu_0}/B_0$, the magnetic Ekman number $\mathrm{Em} = \eta/\Omega R^2$ sets the ratio between the rotation time scale and the magnetic diffusion time scale $\tau_\eta = R^2\eta^{-1}$, and finally the Prandtl number $\mathrm{Pr} = \nu/\kappa$ sets the ratio between the thermal $\tau_\kappa = R^2\kappa^{-1}$ and viscous diffusion time scales. Estimates of these four parameters in Mercury's fluid core can be found in Table \ref{tab:parameters}.
\begin{deluxetable}{lcrl}
\tablecaption{Parameters expected in Mercury's interior.\label{tab:parameters}}
\tablehead{\colhead{Parameter} & \colhead{Symbol} & \colhead{Value} & \colhead{Unit}}
\startdata
Mean rotation rate\tablenotemark{\footnotesize a}                           & $\Omega$          & $1.24\times10^{-6}$   & $\SI{}{\per\second}$              \\
Outer core radius\tablenotemark{\footnotesize b}                            & $R$               & $2015$                & $\SI{}{\kilo\meter}$              \\    
Core kinematic viscosity\tablenotemark{\footnotesize c}                     & $\nu$             & $10^{-6}~$            & $\SI{}{\square\meter\per\second}$ \\
Core magnetic diffusivity\tablenotemark{\footnotesize c}                    & $\eta$            & $1$                   & $\SI{}{\square\meter\per\second}$ \\
Core thermal diffusivity\tablenotemark{\footnotesize c}                     & $\kappa$          & $7\times10^{-6}$      & $\SI{}{\square\meter\per\second}$ \\
Mean outer core density\tablenotemark{\footnotesize b}                      & $\bar{\rho}$      & $7110$                & $\SI{}{\kilogram\per\cubic\meter}$\\
Mean magnetic field near the CMB\tablenotemark{\footnotesize d}             & $B_0$             & $390$                 & $\SI{}{\nano\tesla}$              \\
Magnetic permeability\tablenotemark{\footnotesize c}                        & $\mu_0$           & $4\pi \times 10^{-7}$ & $\SI{}{\newton\per\square\ampere}$\\ 
Libration frequency\tablenotemark{\footnotesize a}                          & $\omega$          & $8.3 \times 10^{-7}$  & $\SI{}{\per\second}$              \\ 
Libration amplitude\tablenotemark{\footnotesize a}                          & $\epsilon$        & $1.9\times 10^{-4}$   & $\SI{}{\radian}$                  \\ 
Equatorial flattening\tablenotemark{\footnotesize e}                        & $\alpha_2$        & $5 \times 10^{-4}$    &                                   \\ 
Moment of inertia of the mantle and crust\tablenotemark{\footnotesize b}    & $C_\mathrm{m+cr}$ & $2.9 \times 10^{35}$  & $\SI{}{\kilogram\square\meter}$   \\ 
\hline
Ekman number                                                                & $\mathrm{Ek}$     & $2 \times 10^{-13}$   &   \\
Magnetic Ekman number                                                       & $\mathrm{Em}$     & $2 \times 10^{-7}$    &   \\
Lehnert number                                                              & $\mathrm{Le}$     & $2 \times 10^{-6}$    &   \\
Prandtl number                                                              & $\mathrm{Pr}$     & $0.1$                 &   \\
Magnetic Prandtl number                                                     & $\mathrm{Pm}$     & $10^{-6}$             &   \\
Elsasser number                                                             & $\Lambda$         & $10^{-5}$             &   \\ 
Permeability ratio between mantle and core\tablenotemark{\footnotesize f}  & $c$          & $\leq 10^{-4}$        &   \\
Conductance ratio between mantle and core\tablenotemark{\footnotesize f}   & $c'$         & $\leq 10^{-4}$        &   \\
\enddata
\tablecomments{The non-dimensional parameters $\mathrm{Ek, Em, Le, Pr, Pm}$ and $\Lambda$ are calculated using the reported physical parameters in the top rows.}
\tablenotetext{a}{\citet[][]{bertone2021deriving}.}
\tablenotetext{b}{Preliminary Reference Mercury Model (PRMM) values taken from \citet[][]{margot2018mercury}.}
\tablenotetext{c}{\citet[][]{wicht2007origin}.}
\tablenotetext{d}{R.m.s. value of the radial magnetic field at the CMB computed from the reconstructed $g_1^0 = \SI{190}{nT}$ component of Mercury's internal magnetic field \citep[][]{anderson2012low}.}
\tablenotetext{e}{\citet[][]{perry2015low}.}
\tablenotetext{f}{Upper bound computed assuming a conducting layer width $\delta \leq \SI{1}{\kilo\meter}$ and equal permeabilities and conductivities in mantle and core, see also Appendix \ref{app:poltor_bound}.}
\end{deluxetable}

The dimensionless \BVfreq $N(r) = \Omega^{-1}(\alpha g_0 ~ dT_0 / dr)^{1/2}$ is a measure of the fluid's stability with respect to convection. The exact shape of this function in Mercury's fluid outer core is unknown but thermal evolution models \citep[see e.g.][]{knibbe2018thermal, knibbe2021modelling} strongly suggest that the top region of the fluid core is stably stratified, i.e. stable against convection with $N^2 > 0$, while the deeper parts are convectively unstable with $N^2 \lesssim 0$. We are particularly interested in the effect of a stably stratified outer layer on the core flow, and following \citet[][]{vidal2015quasi} we write the stratification profile as:
\begin{equation}
N(r) = N_0\sqrt{\frac{1}{2}\left(1 + \tanh\left[\frac{2(r - r_\text{C})}{h}\right]\right)}~. \label{eq:BV_profile}
\end{equation}
This describes a neutrally stratified ($N^2 = 0$) core with radius $r_C$ that smoothly transitions into a stably stratified region at the top of the core with maximum \BVfreq $N_0$, where smoothness parameter $h$ governs the width of the transition region.

\subsection{Spectral method}
We solve Equations \eqref{eq:momentum_nd}-\eqref{eq:incompressible_nd} using a fully spectral decomposition of the variables. Scalar variables such as the temperature field can be written as:
\begin{equation}
\Theta(t, r,\theta,\varphi)=e^{i\omega t}\sum_{\ell=0}^L\sum_{m=-\ell}^\ell\sum_{k=0}^N\Theta_{k,\ell,m}\mathrm{T}_k(r)\mathrm{Y}_\ell^m(\theta,\varphi)~ + \mathrm{c.c.}, \label{eq:ThetaYlm}
\end{equation}
where $\mathrm{T}_k(x)$ are the Chebyshev polynomials of the first kind. The velocity and magnetic field increments are solenoidal ($\mathbf{\nabla\cdot u} = 0 $ and $\mathbf{\nabla\cdot b} = 0 $), and can be decomposed in a poloidal and toroidal vector field as \citep[e.g.][]{chandrasekhar1981hydrodynamics}:
\begin{equation}
\mathbf{u}=\mathbf{\nabla}\times\mathbf{\nabla}\times\left(U\mathbf{r}\right)+\mathbf{\nabla}\times\left(V\mathbf{r}\right)~,\label{eq:poltoru}
\end{equation}
\begin{equation}
\mathbf{b}=\mathbf{\nabla}\times\mathbf{\nabla}\times\left(F\mathbf{r}\right)+\mathbf{\nabla}\times\left(G\mathbf{r}\right)~,\label{eq:poltorb}
\end{equation}
where $U$, $V$, $F$, and $G$, are scalar functions that can be expanded according to Equation \eqref{eq:ThetaYlm}. 

We convert the differential Equations \eqref{eq:momentum_nd}-\eqref{eq:incompressible_nd} subject to the appropriate boundary conditions (Appendix \ref{app:poltor_bound}) into a set of algebraic equations in the complex coefficients $\{\Theta_{k,\ell,m},U_{k,\ell,m},V_{k,\ell,m},F_{k,\ell,m},G_{k,\ell,m}\}$ using a collocation method. To solve the resulting generalised eigenvalue problem, we use the code \texttt{Kore} \citep[][]{triana2022kore}, which implements the efficient spectral discretisation introduced by \citet{olver2013fast} for the radial direction \citep[see][for details]{rekier2018inertial,triana2019coupling}. This very efficient spectral method gives an algebraic representation of the differential equations involving only sparse matrices, leading to a small memory-footprint, which makes it ideally suitable for core flow studies at low Ekman numbers that typically require a large spatial resolution, especially near the boundary. In our case we empirically choose the best resolution for each the Ekman number, varying $L$ and $N$ in Equation \eqref{eq:ThetaYlm} between 400 for the highest and 1300 for the lowest Ekman numbers. Our choice of Chebyshev polynomials for the radial discretisation further allows us to choose a set of collocation points that is dense near both boundaries, increasing for instance the radial resolution in the neighbourhood of the outer boundary.

The numerical results returned by \texttt{Kore} depend on our choice in the dimensionless scalar values $\mathrm{Ek}, \mathrm{Em}, \mathrm{Le},$ and $\mathrm{Pr}$, the stratification parameters $N_0, h$, and $r_\mathrm{C}$, the inner core radius $r_\mathrm{ICB}$, and the type of libration forcing characterised by $m, \omega, \epsilon,$ and $\alpha_2$, as seen in Table \ref{tab:parameters2}. Even though our numerical solver is uniquely qualified to efficiently solve for very small Ekman numbers, the expected Ekman number in Mercury's core $\mathrm{Ek} = 2 \times 10^{-13}$, can still not be reached numerically. Instead we solve the problem for a series of higher, numerically accessible, Ekman numbers ($\mathrm{Ek} \geq 10^{-10}$), see Table \ref{tab:parameters2}, and check for asymptotic behaviour. We compute a value for the magnetic Ekman and Lehnert number, using the following relations between these dimensionless numbers:
\begin{equation}
\mathrm{Em} = \frac{\mathrm{Ek}}{\mathrm{Pm}}~, \hspace{4cm} \mathrm{Le} = \sqrt{\Lambda ~ \mathrm{Em}}~,
\end{equation}
where $\mathrm{Pm} = \nu/\eta$ is the magnetic Prandtl number and $\Lambda = {B_0}^2/(\Omega\eta\mu_0\bar{\rho})$ is the Elsasser number predicted for Mercury's fluid core (see again Table \ref{tab:parameters}). This way any asymptotic behaviour inferred for $\mathrm{Ek} = 2 \times 10^{-13}$ is automatically valid for the expected $\mathrm{Em}$ and $\mathrm{Le}$ in Mercury. Finally, we set the Prandtl number to its expected value, $\mathrm{Pr} = 0.1$.

Regarding the free variables used in the stratification profile, we show in Appendix \ref{app:sensitivity} that the results are most sensitive to the value of the \BVfreq near the CMB. We therefore here fix the other stratification parameters $h=0.1$, $r_\mathrm{C} = 0.7$, $r_\mathrm{ICB} = 0.4$, and vary $N_0$ around $N_0 = 100$, equal to a dimensional value of the squared \BVfreq ${N_0}^2 = 1.5 \times 10^{-8}~\SI{}{\per\square\second}$, which is a typical value calculated for Mercury's thermal evolution. One limitation in the choice of $N_0$ follows from the fact that our radial discretisation scheme makes it numerically expensive to deal with very sharp stratification profiles, with $h/N_0 \ll 1$, which requires denser (less sparse) matrices to maintain accuracy. Accordingly we limit ourselves to the range $N_0 \leq 10^4$, including the value $N_0 = 0$ corresponding to a neutrally stratified liquid core. 

The variables describing the libration forcing are set equal to their observed values given in Table \ref{tab:parameters}. In order to probe the different types of waves excitable in the core, we allow the forcing frequency to vary slightly around that of the forced libration $\omega = 0.67$. An overview of the parameter values used in our numerical computations, can be found in Table \ref{tab:parameters2}. 

\begin{deluxetable}{lcl}
\tablecaption{Parameter values used for the numerical computations\label{tab:parameters2}}
\tablehead{\colhead{Parameter} & \colhead{Symbol} & \colhead{Dimensionless Value}}
\startdata
Ekman number            & $\mathrm{Ek}$     & $[10^{-6},\ldots, 10^{-10}]$              \\
Magnetic Ekman number   & $\mathrm{Em}$     & $10^6 \mathrm{Ek}$                        \\
Lehnert number          & $\mathrm{Le}$     & $10^{-5/2}\mathrm{Em}^{1/2}$              \\
Prandtl number          & $\mathrm{Pr}$     & $0.1$                                     \\ \hline
\BVfreq at the CMB      & $N_0$             & $[0, 10^{-4}, \ldots,  10^{4}]$           \\
Smoothness parameter    & $h$               & $0.1$                                     \\
Inner core radius       & $r_\mathrm{ICB}$  & $0.4$                                     \\
Convective core radius  & $r_\mathrm{C}$    & $0.7$                                     \\ \hline
Forcing component       & $m$               & $[-2, 0, 2]$                              \\
Libration frequency     & $\omega$          & $[0.62, \ldots, 0.72]$                    \\
Libration amplitude     & $\epsilon$        & $1.9 \times 10^{-4}$                      \\
Equatorial flattening   & $\alpha_2$        & $5 \times 10^{-4}$               
\enddata
\end{deluxetable}

\section{Preliminary theoretical considerations} \label{sec:theory}
In this section we consider, from a general point of view, the conditions under which the core angular momentum can change. The angular momentum of the core is equal to:
\begin{equation}
    \mathbf{L}=\int_V\bar{\rho}\mathbf{r}\times(\mathbf{u}+\mathbf{\Omega}\times\mathbf{r})~dV~.
\end{equation}
From Newton's second law, the sum of torques exerted by the mantle on the core must be equal to the rate of change of angular momentum taken in the inertial reference frame, $d\mathbf{L}/dt$. By using Reynolds' theorem, the time derivative with respect to the rotating frame can be expressed as:
\begin{equation}
\partial_t\mathbf{L}=\oint_S\bar{\rho}(\mathbf{v}\cdot\hat{\mathbf{n}})\left[\mathbf{r}\times(\mathbf{u}+\mathbf{\Omega}\times\mathbf{r})\right]~dS +\int_V\bar{\rho}\partial_t\left[\mathbf{r}\times(\mathbf{u}+\mathbf{\Omega}\times\mathbf{r})\right]~dV~, \label{eq:dLdtReynolds}
\end{equation}
where the first term on the right-hand side accounts for the motion of the librating outer boundary with velocity $\mathbf{v}$, and where $\hat{\mathbf{n}}$ is the outward normal vector to the boundary. Upon expliciting the time derivative in the volume integral and inserting the equation of motion \eqref{eq:momentum}, we find after some algebra:
\begin{equation}  
\begin{split}
\partial_t\mathbf{L}+\mathbf{\Omega}\times\mathbf{L}=&-\int_V\mathbf{r}\times\nabla \left(P-\frac{\bar{\rho}}{2}|\Omega\times\mathbf{r}|^2+\bar{\rho}\Phi\right)dV + \int_V\bar{\rho}\alpha \left(\mathbf{r}\times\Theta\mathbf{r}\right)dV+\int_V\bar{\rho}\mathbf{r}\times\nu\mathbf{\nabla}^2\mathbf{u}~dV\\
+\mu_0^{-1}\int_V\mathbf{r}\times\left[(\mathbf{\nabla} \times \mathbf{b})\times\mathbf{B}_0\right]~dV~
+&\oint_S\bar{\rho}(\mathbf{v}\cdot\hat{\mathbf{n}})\left[\mathbf{r}\times(\mathbf{u}+\mathbf{\Omega}\times\mathbf{r})\right]~dS,    \label{eq:torquebalance}
\end{split}
\end{equation}
where we have used $\partial_t\mathbf{r}\equiv\mathbf{u}$. The left-hand side of Equation \eqref{eq:torquebalance} represents the rate of change of angular momentum as measured in the inertial frame \citep[see e.g.][]{tilgner2015rotational}. The right-hand side therefore represents the total torque on the fluid core. The first term representing the sum of torques from the internal pressure, centrifugal force, and external gravity field can be rewritten as a surface integral over the outer boundary. This integral is zero if that boundary is spherical. The second term vanishes identically reflecting the fact that the radial buoyancy force produces no torque. Excluding the surface integral, the total torque on the fluid core therefore reduces to the sum of the viscous and electromagnetic torques, which in the non-dimensional form can be written as:
\begin{equation}
\begin{split}
    \mathbf{\Gamma}&=\mathrm{Ek}\int_V\mathbf{r}\times\mathbf{\nabla}^2\mathbf{u}~dV+\mathrm{Le}^2\int_V\mathbf{r}\times\left[(\mathbf{\nabla} \times \mathbf{b})\times\mathbf{B}_0\right]~dV\nonumber\\
    &=\mathbf{\Gamma}_\nu+\mathbf{\Gamma}_\eta~.\label{eq:coretorque}
    \end{split}
\end{equation}
These torques are expected to be very small in the bulk of the fluid core, since $\mathrm{Le}$ and $\mathrm{Ek}$ are small (see Tab.~\ref{tab:parameters}). This is not necessarily true of the viscous torque since strong gradients in $\mathbf{u}$ within the Ekman boundary layer near the CMB can compensate the smallness of $\mathrm{Ek}$, as can be seen by rewriting $\mathbf{\Gamma}_\nu$ as a surface integral:
\begin{equation}
    \mathbf{\Gamma}_\nu=\mathrm{Ek}\oint_S \mathbf{r \times} \left(\mathbf{\nabla u} + \mathbf{\nabla u}^\intercal\right) \mathbf{\cdot \hat{n}}~dS~,
    \label{eq:vtorq}
\end{equation}
where $\mathbf{\nabla u}^\intercal$ is the transpose of the velocity gradient. Similarly, the electromagnetic torque can be rewritten as an integral over the spherical CMB as \citep{rochester1962geomagnetic}:
\begin{equation}
    \mathbf{\Gamma}_\eta=\mathrm{Le}^2 \oint_S \left(\mathbf{r \times b}\right)\left(\mathbf{B}_0 \cdot \hat{\mathbf{n}}\right)~dS~.
    \label{eq:emtorq}
\end{equation}
The magnetic field of the core must match that of the mantle at the CMB. The above integral must be zero for a perfectly insulating mantle, and remains very small for a mantle of low conductivity. We quantify both $\mathbf{\Gamma}_\nu$ and $\mathbf{\Gamma}_\eta$ for our model in the following section. 

In the limit $\mathrm{Ek}=\mathrm{Le}=0$, Equation \eqref{eq:torquebalance} shows that for a spherical CMB the only change to the angular momentum comes from the motion of the boundary. The situation is more complicated for a non-spherical CMB, in which case the first term on the right-hand side of Equation \eqref{eq:torquebalance} may not be zero. For a triaxial ellipsoid, one can show that this term does in fact vanish to first order in the core ellipticity, provided that the fluid core's rotation does not deviate too much from that of a solid body, in which case the pressure torque exactly balances the torque from gravity and centrifugal force \citep[e.g.][]{vanhoolst2009effect, dumberry2013role}. For more general types of fluid motion, \citet{ivers2017enumeration} demonstrated that the flow can be decomposed as an infinite superposition of \textit{inertial modes}, and that only a pair of these modes have non-zero angular momentum. The first of these is called the \textit{spin-over} mode. Its vorticity is uniform and parallel to the equatorial plane and precesses around the rotation axis at a near-diurnal frequency---hence its name---and produces no torque along the rotation axis. It is therefore irrelevant to libration, but has relevance to precession-nutation \citep{triana2019coupling,rekier2022free}. The other mode has a vorticity parallel to the rotation axis and an exactly zero frequency. Inserting this solution into Equation \eqref{eq:momentum_nd}, and setting the viscosity and magnetic field to zero, we find that the pressure gradient and the Coriolis force are in a state of equilibrium corresponding to the so-called \textit{geostrophic balance} resulting in a net zero torque on the CMB \citep{kuang1997dynamics}. This balance is upset when viscosity or the magnetic field are present. 

One possible way to relax the geostrophic balance constraint is to introduce non-radial density stratification, in which case the second term on the right-hand side of Equation \eqref{eq:torquebalance} no longer vanishes (as $\Theta\mathbf{r}$ gets replaced by a vector term that is no longer parallel to $\mathbf{r}$). \citet{vidal2020acoustic} provided numerical solutions for the inertial modes in the compressible fully stratified triaxial ellipsoid. We computed the angular momentum for these modes, and found that it is negligible for realistic values of the Mach number and that it vanishes entirely in the incompressible (Boussinesq) limit considered here.

\section{Viscous and electromagnetic torques} \label{sec:torque}
In Section~\ref{sec:theory} we identified the viscous and electromagnetic torques as the primary cause of change in the angular momentum of Mercury's fluid core. By virtue of Newton's third law, these torques must produce a reaction equal in magnitude and opposite in sign on the mantle contributing to the azimuthal libration. 
\begin{figure}
\centering
\plotone{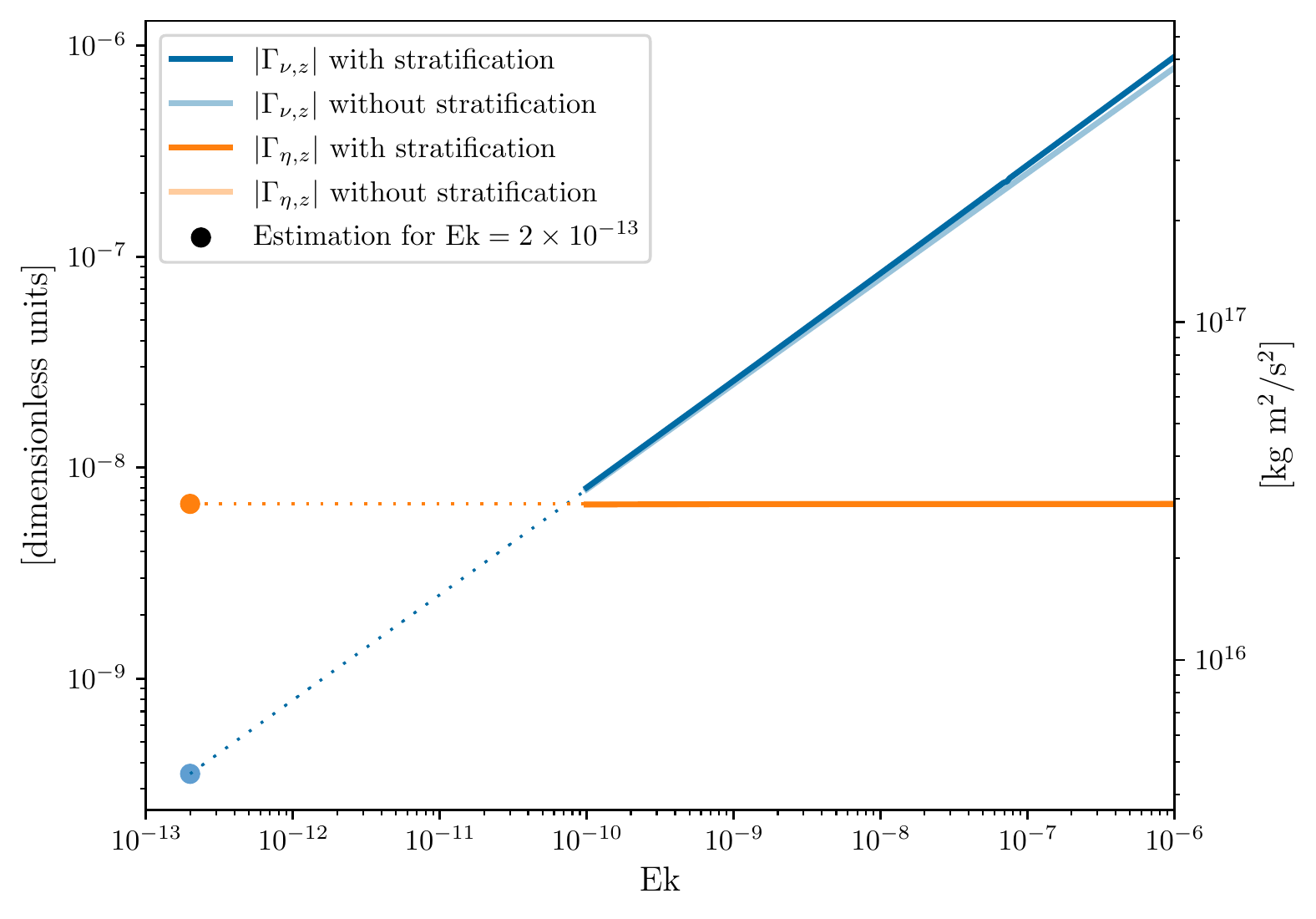}
\caption{Axial viscous $|\Gamma_\nu, z|$, in blue, and electromagnetic torque $|\Gamma_\eta, z|$ in orange, as a function of the Ekman number $\mathrm{Ek}$, applied on the mantle by core motions resulting from the $m=0$ component of libration. Dimensionless values can be read from the left axis and dimensional values from the right axis. The lighter lines corresponds to the case with no background stratification $N(r) = 0$ for $0.4 = r_\mathrm{ICB} \leq r \leq  r_\mathrm{CMB}$ and the darker curve corresponds to the stratified case, Equation \eqref{eq:BV_profile}, with $N_0 = 100, h=0.1, r_\mathrm{C} = 0.4$, and $r_\mathrm{ICB} = 0.7$. Extrapolation to the torques expected in Mercury's core, represented by the filled circles, is performed along the dotted lines.}
\label{fig:torque}
\end{figure}
Figure \ref{fig:torque} shows the total axial viscous and axial electromagnetic torque on the mantle caused by the librationally induced core flow. Only the $m=0$ forcing component contributes to the torques as the integration of any spherical harmonic with $m \neq 0$ over the spherical surface returns zero. The viscous torque in Figure \ref{fig:torque} is slightly higher with stratification than without. The differences are small, however, and decrease with decreasing Ekman number. Extrapolating these differences to the expected Ekman number in Mercury's core, $\mathrm{Ek} = 2\times 10^{-13}$, we find that the additional viscous torque due to the stratification is less than one percent of the total viscous torque in a core with no stratification, meaning that the stratified layer in Mercury's core has little to no influence on the viscous torque applied to the mantle. We will address the reason for this small influence in Section \ref{sec:m0} where we discuss the  differences in core flow between the cases with and without a stably stratified layer. 

In the neutrally stratified core the axial viscous torque is proportional to the square root of the Ekman number:
\begin{equation}
|\Gamma_{\nu,z}| = 0.00077~ \mathrm{Ek}^{1/2}~, \label{eq:func_vtorq}
\end{equation}
which implies that the viscous torque decreases with the square root of the Ekman number. Such a scaling points to a relation between the viscous torque and the thickness of the Ekman boundary layer, also proportional to $\sqrt{\mathrm{Ek}}$. This is consistent with the results of \citet[][Equation 10]{triana2019coupling} and \citet[][Equation 22]{cebron2019precessing} who identify the same scaling for the power dissipated at the boundary, which is in itself proportional to the axial torque, the two quantities being related by a common coupling constant \citep[][]{triana2021viscous, cebron2019precessing,deleplace2006viscomagnetic}. By assuming that the relation \eqref{eq:func_vtorq} persists for all smaller values of the Ekman number, the axial viscous torque applied by Mercury's outer core would be only $1.3 \times 10^{14} ~\SI{}{\newton\meter}$, less than $\SI{0.001}{\percent}$ of the total torque needed to drive libration $|\Gamma_{\mathrm{tot},z}|$, which we estimate as:
\begin{equation}
|\Gamma_{\mathrm{tot},z}| = C_\mathrm{m+cr}|\ddot{\phi}| = C_\mathrm{m+cr}\omega^2 \epsilon = 3.7 \times 10^{19}~ \SI{}{\newton \meter}~,
\end{equation}
where we use the values of the moment of inertia of the mantle plus crust $C_\mathrm{m+cr}$, and the angular acceleration $|\ddot{\phi}|$ of the mantle caused by libration (Table \ref{tab:parameters}). Since the change in total torque due to viscous effects implies rotation variations far below the current observational precision \citep{bertone2021deriving} of Mercury's libration amplitude, we conclude that, barring any substantial additional torque due to non-linear instabilities, the axial viscous torque at the core-mantle boundary can safely be neglected in studies of the planet's libration, confirming and strengthening the conclusion of \citet[][]{peale2002procedure} which was based on estimates of the viscous relaxation time scale of relative motion between the core and the mantle.

\begin{figure}
\centering
\epsscale{0.5}
\plotone{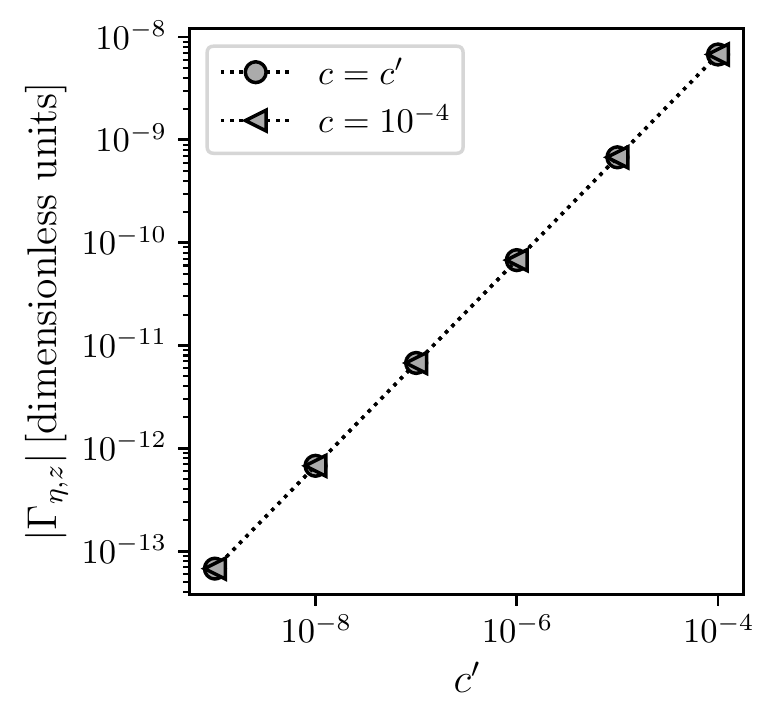}
\caption{Electromagnetic torque at the outer boundary as a function of the thin wall parameters $c$ and $c'$. Triangles mark cases where $c$ is fixed to $10^{-4}$ and only $c^\prime$ is varied, while circles point to cases where $c$ is equal to $c^\prime$ and varied. The results presented here are computed for $\mathrm{Ek} = 10^{-6}$, although they are identical to computations with a different Ekman number.}
\label{fig:cc}
\end{figure}

The total electromagnetic torque remains approximately equal for all parameter values considered ($10^{-10} \leq \mathrm{Ek} \leq 10^{-6}$ and $N_0 = \{0, 100\}$). This indicates that neither the stratification nor the Ekman number can affect the torque magnitude, which is in all cases very small, only about $6.7\times 10^{-9}$ or $4.6\times 10^{15}~\SI{}{\newton\meter}$. The only variable that has any influence on the magnitude of the electromagnetic torque is $c^\prime$, one of the parameters used in the thin wall boundary condition at the CMB. Figure \ref{fig:cc} shows that the torque reaches its maximum when $c'$ is maximal: i.e. at the limit of validity of the thin wall approximation ($c' = 10^{-4}$, see Equation \eqref{eq:thinwall2}), which is the case displayed in Figure \ref{fig:torque} ($c = c' = 10^{-4}$). Assuming that the behaviour shown in Figure \ref{fig:torque} does not change for lower Ekman numbers, the linear electromagnetic torque is at most $\SI{0.013}{\percent}$ of the total torque, indicating that even though at Mercurial conditions it is larger than the viscous torque, it also doesn't significantly affect the observed libration amplitude and accordingly does not need to be considered in studies of Mercury's libration. The inclusion of turbulent effects might, as alluded to above, change this picture. \citet[]{cebron2019precessing} showed for instance in the precessing lunar core that the viscous dissipation, related to the viscous torque, could increase by an order of $10^4$ (enough to become comparable to the observed libration). However this result is only valid for some extreme parameter cases, that are unrealistic for Mercury's core.

\section{Influence of the stratified layer on the librationally induced core flow} \label{sec:results}

\subsection{The core flow excited by the axial libration forcing} \label{sec:m0}
\begin{figure}
\centering
\plotone{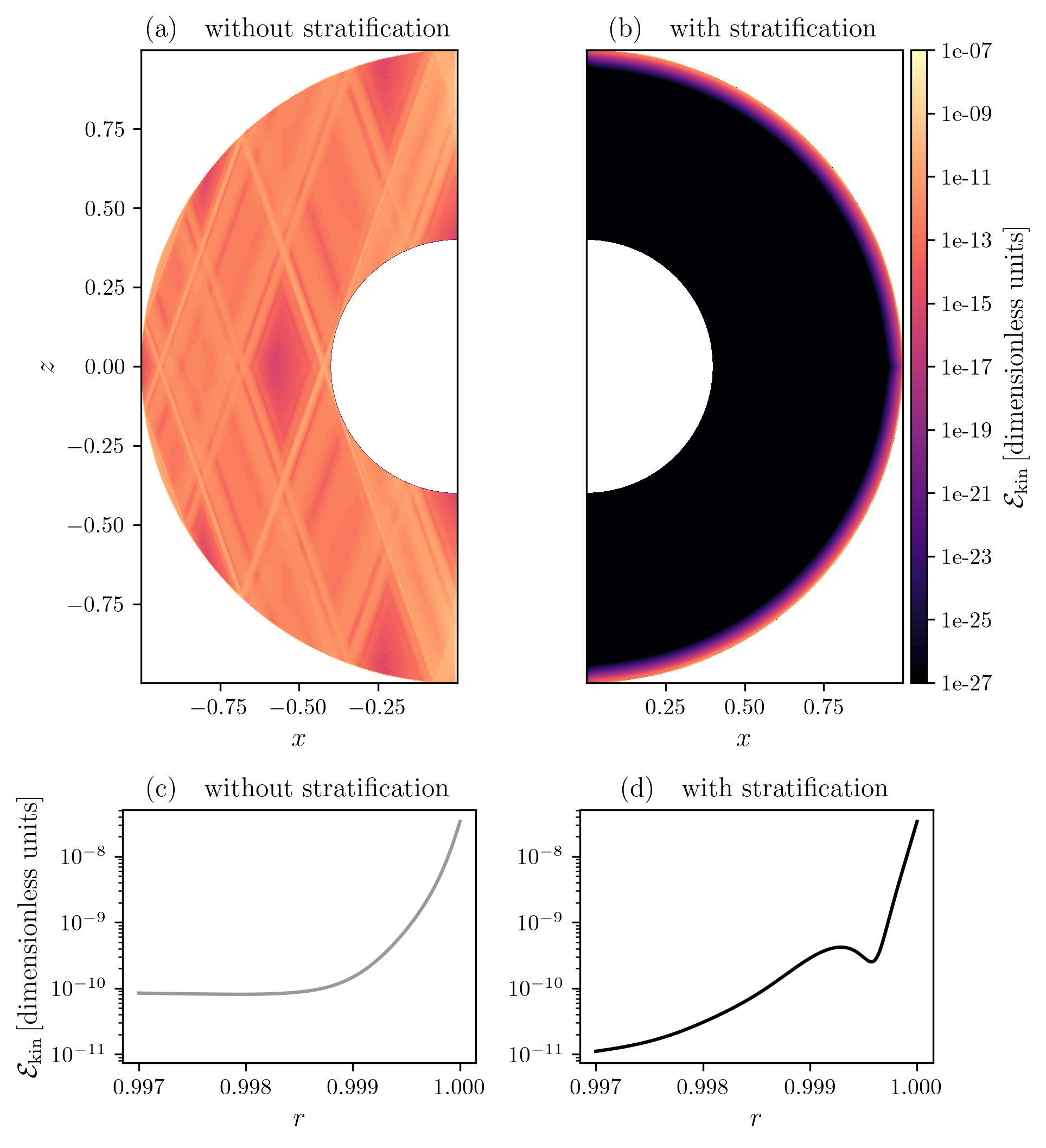}
\caption{Kinetic energy density of the flow, $\mathrm{Ek} = 10^{-8}$, in response to the ($m=0$, $\omega=0.67$) libration forcing in a non-stratified core with $r_\mathrm{ICB} = 0.4$ and $N_0 = 0$ (left column) and in a stratified core with $r_\mathrm{ICB} = 0.4$, $r_\mathrm{C} = 0.7$, $h=0.1$, and $N_0 = 100$ (right column). The top row displays meridonial cuts ($\phi = 0$) of the kinetic energy density while the bottom row shows the total kinetic energy density as a function of the radial coordinate.}
\label{fig:damped_mode}
\end{figure}

\subsubsection{Stratification strongly suppresses radial flow motions and resonance effects with (gravito-)inertial modes}
Although the presence of a stably stratified layer has little effect on the torques acting on the CMB, stratification can substantially alter the librationally induced flow generated near the outer boundary. For this result and the results that follow we have neglected the Lorentz force ($\mathrm{Le}^2\left[(\mathbf{\nabla} \times \mathbf{b} ) \times \mathbf{B}_0\right]$) in the momentum equation, because as shown in Figure \ref{fig:testB} in Appendix \ref{app:sensitivity} the back reaction of the Lorentz force on the flow is negligible due to the weakness of our background magnetic field. Figure \ref{fig:damped_mode} shows the kinetic energy density of the core flow excited by the $m=0$ component of Mercury's libration both without ($N_0 = 0$) and with stratification ($N_0 = 100$). At first glance it might seem that the kinetic energy of the non-stratified core is several orders of magnitude larger than that of the stratified core, but the total kinetic energy of the outer core actually only differs one order of magnitude: about $1.04 \times 10^{-11}$ in the non-stratified core and $3.02 \times 10^{-12}$ in the core with a top stratified layer (dimensionless units). This is because the total kinetic energy is almost entirely determined by the kinetic energy density maximum near the boundary, which is similar in both cases, see Figure \ref{fig:damped_mode}(c)-(d). In the deeper parts of the core the kinetic energy density is orders of magnitude smaller than the maximum value but there the differences between the two cases can be very large. Without stratification, the flow can be radially transmitted throughout the entire core, and, as Figure \ref{fig:damped_mode}(a) shows, the kinetic energy reaches a local maximum in internal shear layers, as expected for inertial modes \citep[see e.g.][]{tilgner1999driven}. With stratification the excited flow follows the azimuthal motion of the outer boundary and is confined to the region near the boundary, and the radial flow decays very quickly towards deeper parts of the core, Figure \ref{fig:damped_mode}(d). The slightly larger velocity gradient near the CMB in the stratified case explains why the viscous torque is slightly larger in that case, cf. Equation \eqref{eq:vtorq}.

\begin{figure}
\centering
\epsscale{1.09375}
\plotone{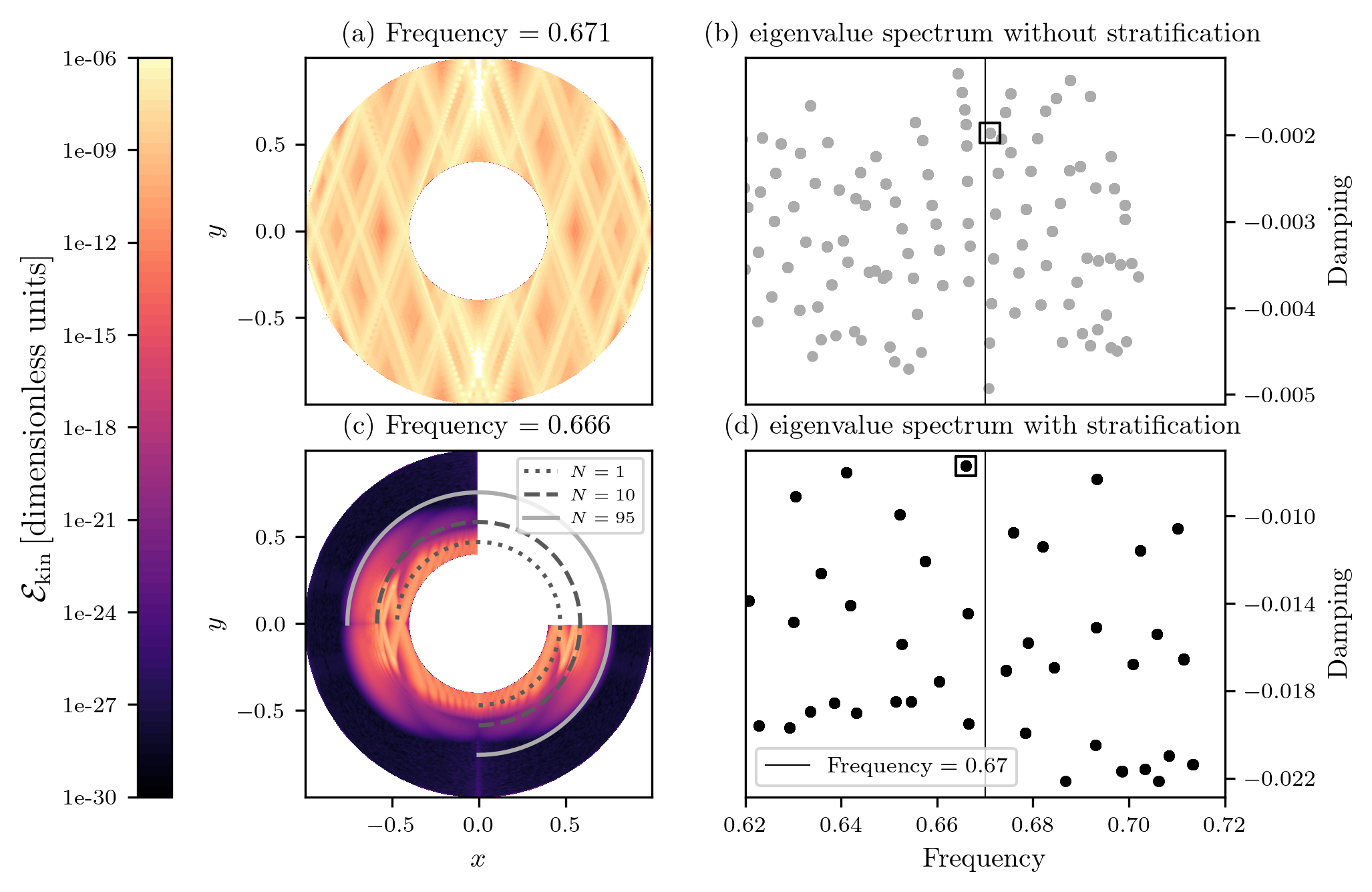}
\caption{Eigenvalue spectra of the unforced problem ($\epsilon = 0$) without stratification ($N_0 = 0$, upper panels) and with stratification ($N_0 = 100$, lower panels). Figures (b) and (d) display the damping, $\mathrm{Im}[\omega]$, versus the frequency, $\mathrm{Re}[\omega]$, of the least-damped eigensolutions near the forcing frequency ($\omega = 0.67$) of Mercury's longitudinal libration, denoted by the vertical black line. The eigensolution with frequency closest to this libration frequency is in both panels marked with a black square and meridonial cuts of their corresponding kinetic energy density are shown on the left in panels (a) and (c). }
\label{fig:spectrum}
\end{figure}

The reason for the flow differences between the stratified and the non-stratified case can be understood by looking at the eigenvalue spectrum, the set of complex frequencies of the free modes solutions to our system of Equations \eqref{eq:momentum_nd}-\eqref{eq:incompressible_nd} with forcing amplitude $\epsilon$ equal to zero. Without stratification and a magnetic field the solutions to this problem are inertial modes. Figure \ref{fig:spectrum}(b) shows frequency and damping of some of these inertial modes with frequencies ranging from $0.62$ to $0.72$ containing the frequency $\omega=0.67$ of Mercury's longitudinal libration in the forced problem. A meridonial cut of the kinetic energy density of the inertial mode closest to this forcing frequency is given in Figure \ref{fig:spectrum}(a). As is characteristic for inertial modes, this mode exhibits internal shear layers, regions of intense shear (with high velocity gradients) that in Figure \ref{fig:spectrum}(a) can be identified as straight lines along which the kinetic energy is maximal. The pattern of these internal shear layers is very similar to the forced flow structure in Figure \ref{fig:damped_mode}(a), indicating that in the neutrally stratified core the libration forcing at the boundary excites a flow that has the same spatial structure as a nearby inertial mode. 

Strong stratification significantly changes the eigenvalue spectrum (Figure \ref{fig:spectrum}(d)) and the flow structure of the eigenmodes (Figure \ref{fig:spectrum}(c)). In the deep interior where the \BVfreq is below the rotation frequency ($N<1$ in dimensionless units), the flow behaves like an inertial mode with internal shear layers present at all latitudes. Where the stratification is stronger, internal shear layers are also present but they are trapped in a region near the equator, which is typical of gravito-inertial modes with $0 < \omega < 2N$ \citep[see e.g.][]{dintrans1999gravito}. In the outer regions of the core, where the stratification is large compared to rotation ($N \gg 10$), shear layers are absent and the flow behaves like a gravity mode, whose propagation direction depends on the magnitude of the \BVfreq:
\begin{equation}
\omega^2 = N^2 \cos^2\gamma
\end{equation}
where $\gamma$ is the angle between the group velocity of the wave and $\mathbf{\hat{r}}$. For very high $N$, $N \gg \omega = 0.67$ implies $\cos\gamma \ll 1$ and the group velocity and particle motion of the gravity modes is almost tangential, so that most of the radial energy transfer and motion is suppressed. Consequently the energy from the libration forcing at the boundary cannot be transmitted by radial motions towards the bulk of the core, and does not excite a flow with the same spatial structure as in Figure \ref{fig:spectrum}(c). Instead, the extent of the librationally induced flow is limited to the immediate vicinity of the boundary where it is driven (Figure \ref{fig:damped_mode}(b)). This suppression of radial energy is thus what prevents the flow from being similar to the flow of the gravito-inertial mode that is closest in frequency to the forcing frequency. Note that the difference in flow structure of the forced response cannot be explained by the differences in damping, as the damping only changes by a factor of 4 between the cases with and without stratification.

\begin{figure}
\centering
\epsscale{1.15625}
\plotone{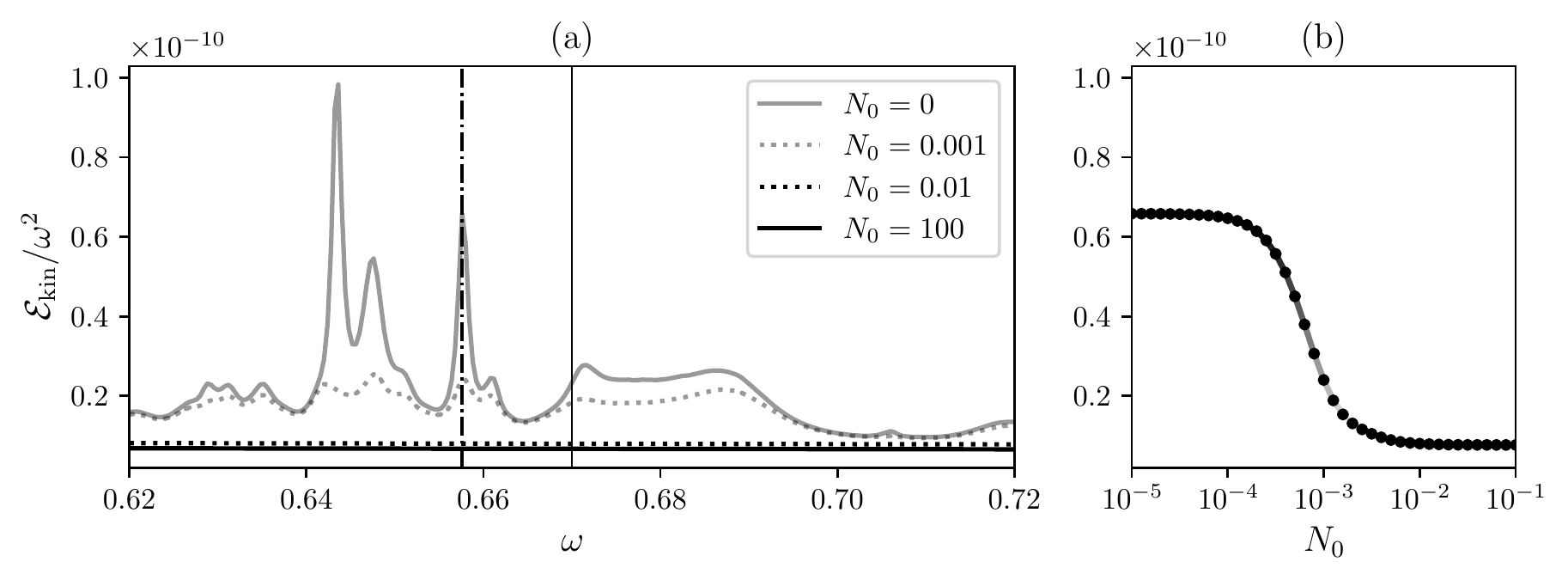}
\caption{Kinetic energy density inside the core for the $m=0$ libration forcing as a function of (a) the forcing frequency $\omega$ for different profiles of the \BVfreq at the CMB and (b) the \BVfreq $N_0$ at peak forcing frequency $\omega = 0.6576$ (denoted by the dash-dotted vertical line). The black vertical line coincides with the observed libration frequency $\omega=0.67$ in Mercury and the remaining model parameters are $E = 10^{-8}$, $r_\mathrm{C} = 0.7$, $h=0.1$ and $r_\mathrm{ICB} = 0.4$.}
\label{fig:resonances}
\end{figure}

Figure \ref{fig:resonances} shows that this suppression of radial motion already happens for weak stratification starting from $N_0 \geq 0.01$ for which the flow behaviour is expected to be inertial as mentioned in the above paragraph. In Figure \ref{fig:resonances}(a) peaks of the kinetic energy profile correspond to resonances with gravito-inertial modes and are related to the internal shear layers in the bulk of the core. Around $N_0 = 0.01$ (Figure \ref{fig:resonances}(b)) the peaks disappear, indicating that suppression of radial motion is sufficient to prevent inertial mode-like behaviour in the bulk of the core.  Figure 5(b) shows that this transition is sharp, suggesting that the suppression of the internal shear layer strength by the stratified top layer is very efficient, although further study into this phenomenon is needed. 

\subsection{The core flow excited by the radial libration forcing}\label{sec:m2}
\subsubsection{Stratification induces a strong horizontal flow near the outer boundary} 
\begin{figure}
\epsscale{1.09375}
\centering
\plotone{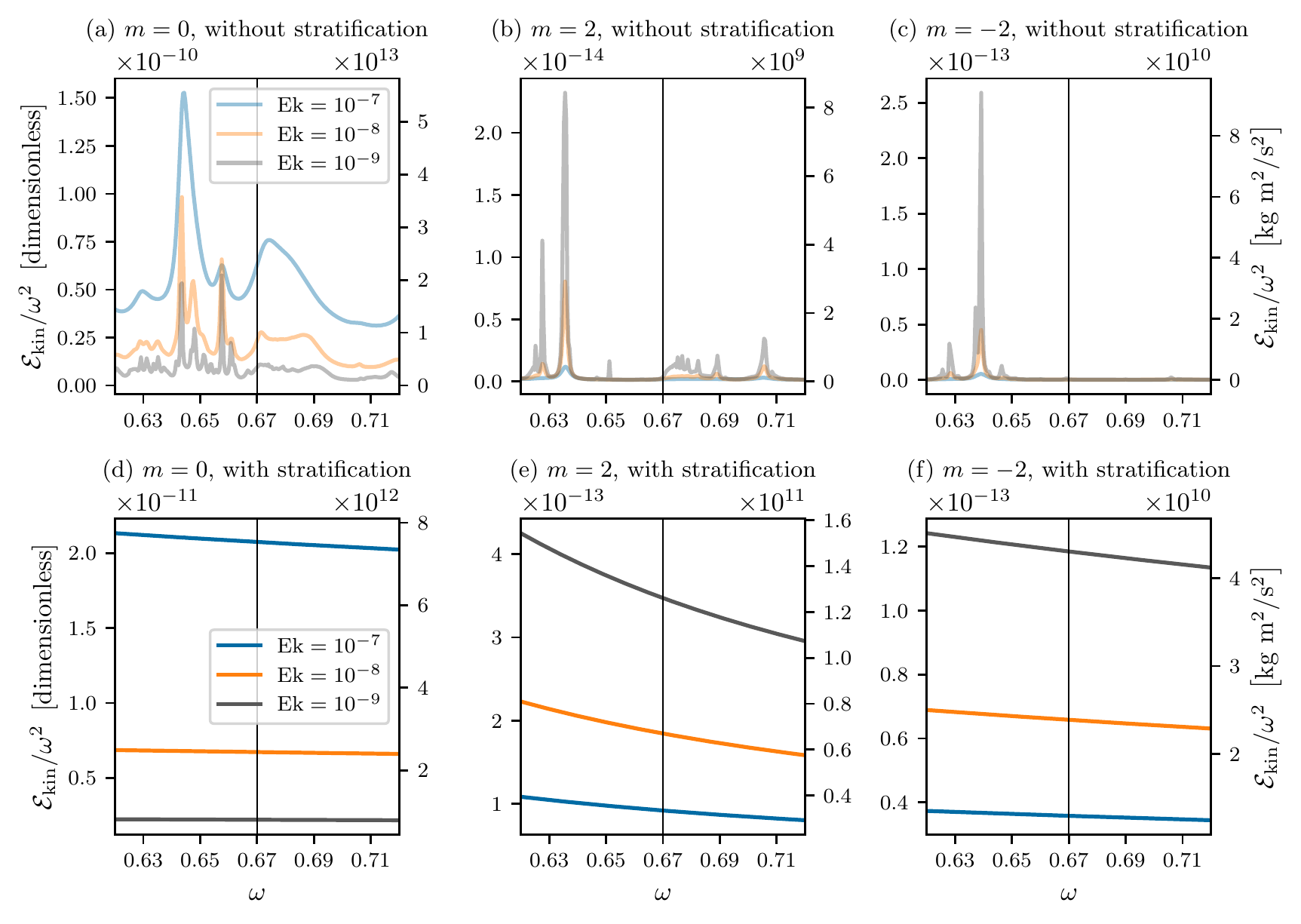}
\caption{Total kinetic energy density of the librationally induced core flow as a function of the forcing frequency $\omega$, near Mercury's observed libration frequency, $\omega=0.67$, indicated in each plot by the black vertical line. The top row gives the core response to the three types of libration forcing, $m = 0, \pm 2$ in a neutrally stratified core ($N_0 = 0$, $r_\mathrm{ICB}=0.4$), and the bottom row shows the response in a core with an outer stably stratified layer ($N_0 = 100$, $r_\mathrm{ICB}=0.4$, $r_\mathrm{C} = 0.7$, and $h=0.1$). Different coloured lines correspond to different Ekman numbers: $\mathrm{Ek} = 10^{-7}$ (blue), $\mathrm{Ek} = 10^{-8}$ (orange), and $\mathrm{Ek} = 10^{-9}$ (dark grey).}
\label{fig:kinetic_energy}
\end{figure}
The suppression of radial motions by the stratified layer is also evident in the flow induced by the $m=\pm 2$ components of libration (Figure \ref{fig:kinetic_energy}): for all forcing types ($m=0, \pm2$) peaks can be observed in the kinetic energy profile of the flow in the core without stratification, but these disappear when stratification is introduced. There are fewer peaks for higher Ekman numbers as the eigenvalue spectrum becomes less dense with viscosity \citep[see e.g.][]{rieutord1997inertial} and due to the increased damping, the peaks become larger and merge into each other.

For the Ekman numbers considered in Figure \ref{fig:kinetic_energy} the core flow induced by the $m=\pm 2$ components of the libration forcing is smaller than the flow from the $m=0$ component, but this is not necessarily the case for lower Ekman numbers. The kinetic energy of the fluid flow decreases with decreasing Ekman number for the axial forcing component, while it increases with decreasing Ekman number for the radial forcing components. The opposite behaviour is related to the different mechanisms of angular momentum transfer between the $m=0$ and the $m=\pm 2$ forcings. For $m=0$, within the ranges of Ekman numbers considered, the viscous torque is the main transfer mechanism. Since that torque decreases with decreasing $\mathrm{Ek}$ (Figure \ref{fig:torque}) the kinetic energy is smaller for smaller $\mathrm{Ek}$. In the $m=\pm 2$ case the radial in- and outflow from the boundary transfers the motion of the mantle to the core. This mechanism does not depend on viscosity and will therefore not change when the Ekman number decreases, but the excited fluid motions will be less damped by viscous forces for lower Ekman numbers resulting in an increase of fluid kinetic energy. 

\begin{figure}
\centering
\epsscale{1.25}
\plotone{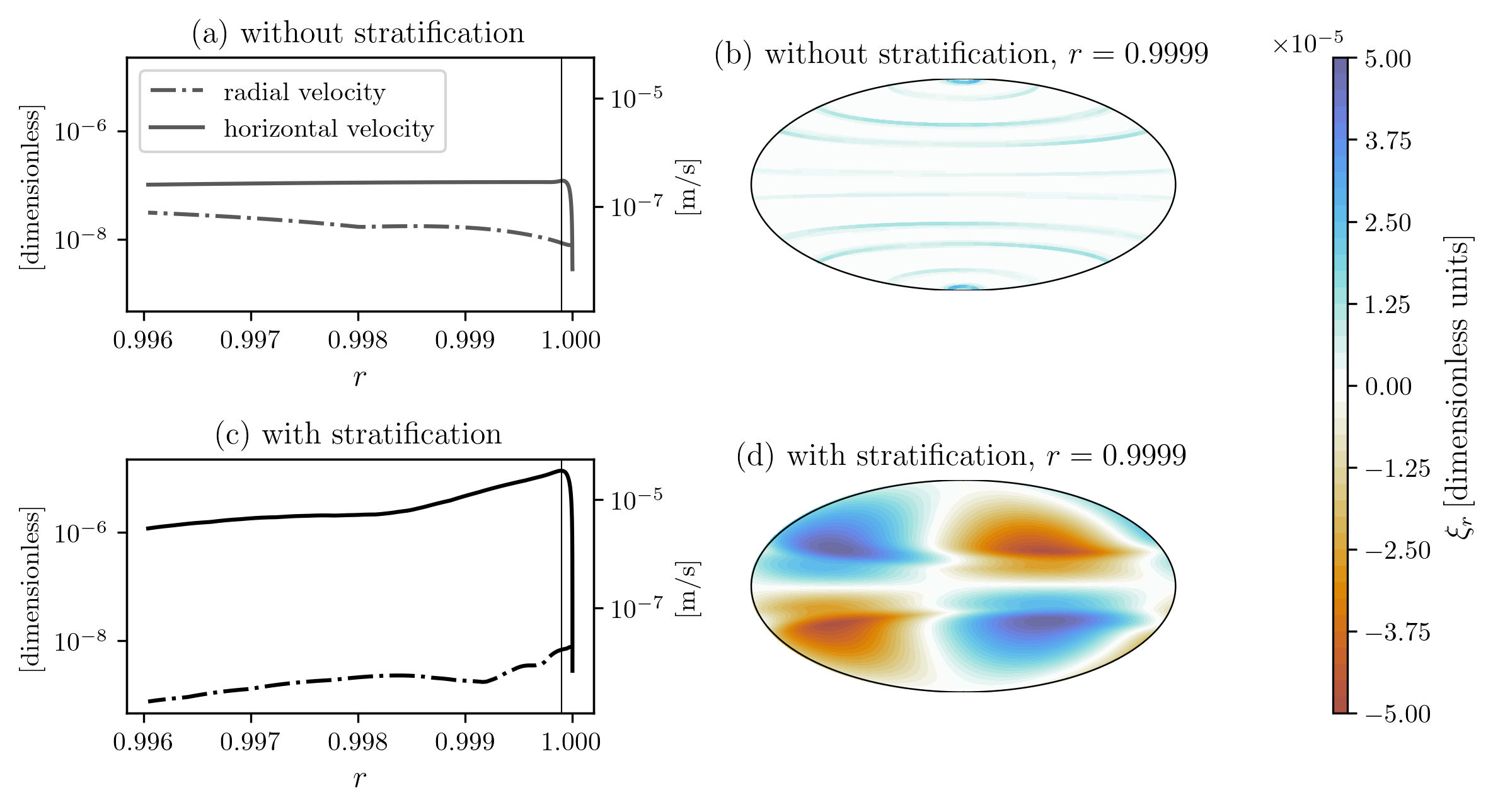}
\caption{Structure of the core flow in response to the $m=2$ libration forcing for $\mathrm{Ek} = 10^{-9}$, with on the left the radial ($u_r$) and horizontal ($({u_\theta}^2 + {u_\phi}^2)^{1/2}$) as a function of the radial coordinate near the CMB and on the right the  radial vorticity $\xi_r = \hat{\mathbf{r}} \mathbf{\cdot \nabla \times u}$ at the (local) maximum of horizontal kinetic energy, that is indicated by the vertical lines in the left figures. The upper panels correspond to the fluid motions in a neutrally stratified core ($N_0 = 0$) and the bottom panels to the motions in a core with a stably stratified outer layer with $N_0 = 100$.}
\label{fig:flow_profile}
\end{figure}

The increase in kinetic energy is particularly significant when the top of the core is stably stratified, in which case we again do not observe any resonant effects with gravito-inertial modes, (Figure \ref{fig:kinetic_energy}(e),(f)) leading to a large horizontal flow very close to the outer boundary, where the fluid reaches its maximum kinetic energy around $r \approx 0.9999$.\footnote{Note that due to the for this case high radial resolution $N = 796$ and our advantageous choice of collocation points, the radial points near the boundary are approximately spaced with $10^{-5}$, allowing us to find values like this.} This radius is just a bit below the Ekman layer that approximately extends down to $r = 0.999968$. Stratification causes the purely radial flow forced by the $m=\pm 2$ components from the boundary to decay very quickly towards deeper parts of the core. By virtue of the incompressibility condition:
\begin{equation}
\frac{1}{r^2}\frac{\partial (r^2 u_r) }{\partial r} = - \nabla^2 \mathbf{u}_\mathrm{tan}    
\end{equation}
this decrease in radial velocity $u_r$ is compensated by a strong increase in horizontal velocity $\mathbf{u}_\mathrm{tan} = u_\theta \boldsymbol{\hat{\theta}} + u_\phi \boldsymbol{\hat{\phi}}$ (Figure \ref{fig:flow_profile}(c)). This horizontal flow component can become much faster compared to both the radial and horizontal flow components without stratification (Figure \ref{fig:flow_profile}(a)) as the horizontal flow here is confined to a much smaller region. Our calculations further reveal that the horizontal flow structure is characterised by four large horizontal vortices (Figure \ref{fig:flow_profile}(d)), reflecting the $\mathrm{Y}_2^2$ shape of the $m=\pm 2$ forcing. Without stratification, the radial flow structure is not suppressed as much (Figure \ref{fig:flow_profile}(a)), and the horizontal flow does not exhibit the same structure (Figure \ref{fig:flow_profile}(b)). 

\subsubsection{Stratification increases the Reynolds number near the outer boundary}
As mentioned above our system of Equations \eqref{eq:momentum_nd}-\eqref{eq:incompressible_nd} is linear and only valid for small perturbations in the velocity field, when the fluid flow is laminar, i.e. characterised by smooth and constant fluid motions. An important parameter to determine whether or not the flow is laminar is the Reynolds number, the dimensionless ratio between inertial and viscous forces, defined as
\begin{equation}
\mathrm{Re} = \frac{\mathcal{U}\mathcal{L}}{\nu}~, \label{eq:reynolds}
\end{equation}
where $\mathcal{U}$ and $\mathcal{L}$ are typical velocity and length scales for the flow motion, respectively. At low Reynolds numbers the fluid is predominantly laminar and the linear system of equations \eqref{eq:momentum}-\eqref{eq:incompressible} is valid, while at high Reynolds numbers the fluid will be turbulent, and non-linear interactions need to be considered. A precise threshold value between the laminar and turbulent flow regimes depends on the exact nature of the problem but we expect flows with $\mathrm{Re} \ll 10$ to be laminar and flows with $\mathrm{Re} \gg 10^{3}$ to be fully turbulent, while some turbulent effects might already be observed above $\mathrm{Re} \gtrsim 140$ in a neutrally stratified core \citep[see e.g.][]{noir2009experimental, calkins2010axisymmetric}. Without the enhanced horizontal flows due to stratification (Figure \ref{fig:flow_profile}(c)-(d)), the Reynolds number can be estimated using the velocity of the librating boundary, $r_\mathrm{CMB}\epsilon\omega$, as the typical velocity scale and the width of the Ekman boundary layer, $(r_\mathrm{CMB} - r_\mathrm{ICB})\sqrt{Ek}$, as the typical length scale. Doing so results in $\mathrm{Re} \leq 7$ for our numerical results and $\mathrm{Re} \approx 172$ in Mercury's fluid outer core, the latter being larger than the threshold value of $\boldsymbol{140}$ defined just above. This means that the nonlinear effects, that we do not consider in this study, for reasons outlined in Sections \ref{sec:intro} and \ref{sec:equations}, do not influence the numerical results presented in this and the previous Section but might as discussed have some influence on the extrapolated values for Mercury in Figure \ref{fig:torque}. On the other hand, similar values for the Reynolds number in a precessing sphere \citep[][]{cebron2019precessing} only increase the viscous dissipation and by extension the viscous torque by a factor of 0.2, whereas an increase of $10^4$ is needed for the viscous and electromagnetic torques to affect the observed libration amplitude.

For the strong horizontal flow (Figure \ref{fig:flow_profile}(c)-(d)), the maximum horizontal velocity near the CMB determines the velocity scale of the fluid motion and its corresponding location, measured as the distance to the outer boundary, determines the length scale. The associated Reynolds number is computed as:
\begin{equation}
\mathrm{Re} = \frac{\max{(\mathbf{u}_\mathrm{tan})}\left(r_\mathrm{CMB} - \max_r{(\mathbf{u}_\mathrm{tan})}\right)}{\mathrm{Ek}}~, \label{eq:comp_reynolds}
\end{equation}
where $\max{(\mathbf{u}_\mathrm{tan})} = ({u_\theta}^2 + {u_\phi}^2)^{1/2}$ is the maximum horizontal velocity and $\max_r{(\mathbf{u}_\mathrm{tan})}$ is the radius where the flow reaches this maximum. 

\begin{figure}
\centering
\plotone{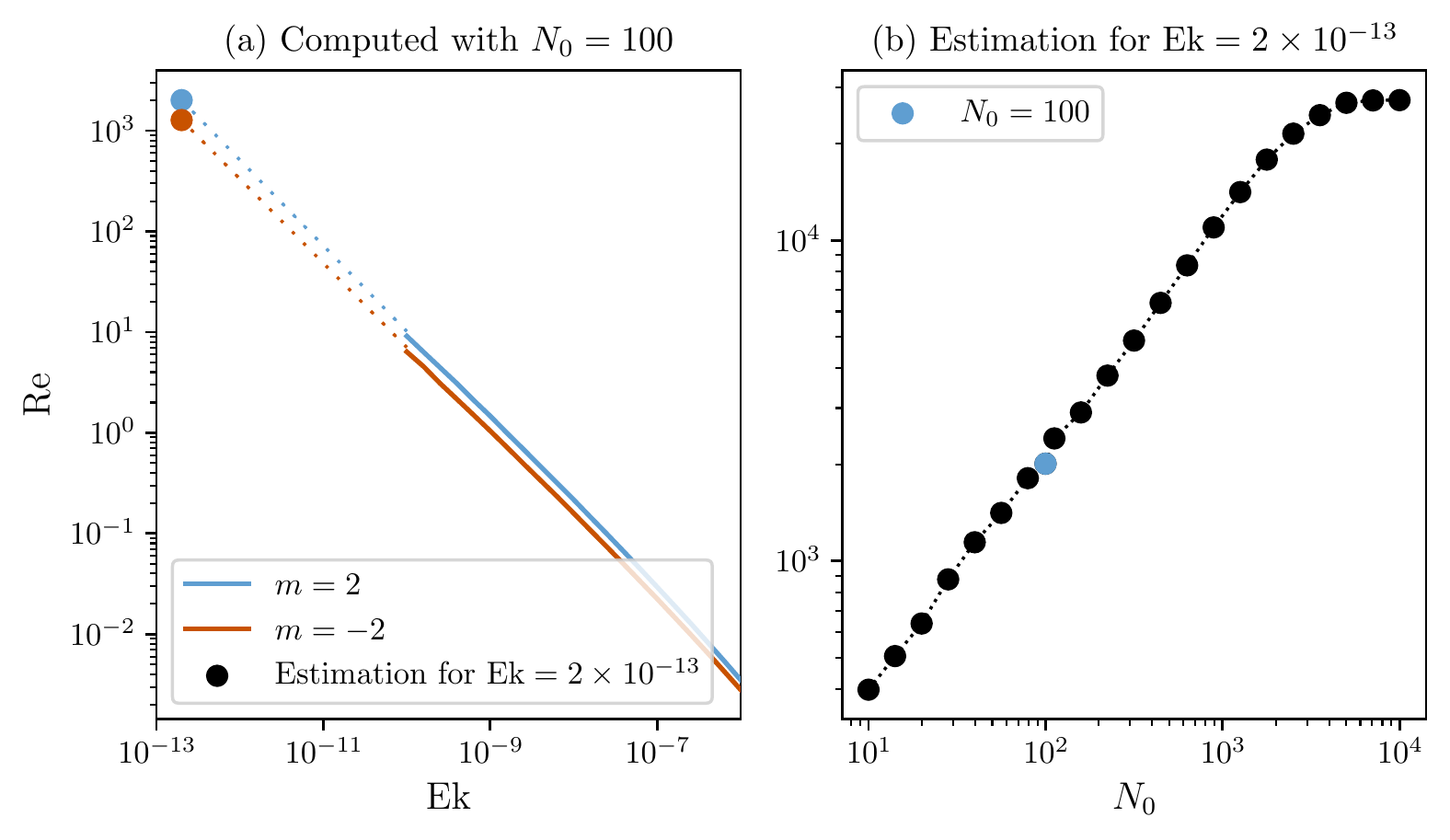}
\caption{(a): Example of the Reynolds number near the core-mantle boundary in function of the Ekman number when the core is stably stratified with stratification parameters $N_0 = 100$, $r_\mathrm{ICB} = 0.4$, $r_\mathrm{C} = 0.7$, and $h = 0.1$ and the mantle is forced by the radial components of libration, $m=2$ (blue) en $m=-2$ (dark red). The expected value of the Reynolds number at Mercurial conditions, $\mathrm{Ek} = 2 \times 10^{-13}$, assuming a linear relationship between $\log{(\mathrm{Re})}$ and $\log{(\mathrm{Ek})}$ is indicated with a star. (b): Extrapolated Reynolds numbers of the $m=2$ core flow response near Mercury's core-mantle boundary for different stratification strengths $N_0$. The blue star represents the Reynolds number estimated from the left figure. The slightly irregular behaviour for $10 \lesssim N_0 \lesssim 10^{2.5}$ is an artefact of the numerical method not being precise enough to determine the location of maximum horizontal velocity when the peak of the horizontal flow profile is less pronounced.}
\label{fig:reynolds}
\end{figure}

We have computed the Reynolds numbers for the radially forced fluid flow assuming a top stably stratified layer with $N_0$ between 10 and 10000 (Figure \ref{fig:reynolds}). We don't consider \BV frequencies below $N_0 = 10$ because the stratification is then too weak to clearly define a peak in the horizontal flow velocity near the boundary. As expected from definition \eqref{eq:reynolds} and \eqref{eq:comp_reynolds}, the Reynolds number increases with decreasing Ekman number (Figure \ref{fig:reynolds}(a)) and the relationship follows a power law $\mathrm{Re} \propto \mathrm{Ek}^a$. Assuming the power law is valid down to the expected value of the Ekman number in Mercury's fluid core, i.e. is mostly unaffected by any nonlinear interactions, we estimate the Reynolds number in Mercury's core to be $\mathrm{Re} \approx 2012$ for $m=2$ and $\mathrm{Re} \approx 1275$ for $m=-2$ for an outer stable layer with $N_0=100$, much larger than our previous estimate of $Re \approx 172$ without stratification. This suggests that stratification increases the Reynolds number near the boundary, and for realistic parameter values of Mercury we would probably observe even more non-linear phenomena than proposed by \citet[][]{noir2009experimental}, although further non-linear studies would be needed to confirm this.

The Reynolds number also increases with increasing \BVfreq (Figure \ref{fig:reynolds}(b)). For all values of the stratification strength $N_0>10$, the Reynolds number is near or above $1000$, meaning that the flow near Mercury's core-mantle boundary is most likely turbulent, unless the \BVfreq is significantly smaller. For $N_0 \gtrsim 10^{3.5}$ the increase in Reynolds number seems almost negligible, teasing the possibility of an asymptotic limit when the top of the horizontal jet becomes closer and closer (but never actually crosses into) the Ekman boundary layer. 

\subsubsection{Stratification induces a non-axisymmetric magnetic field structure}

\begin{figure}
\epsscale{1.15625}
\centering
\plotone{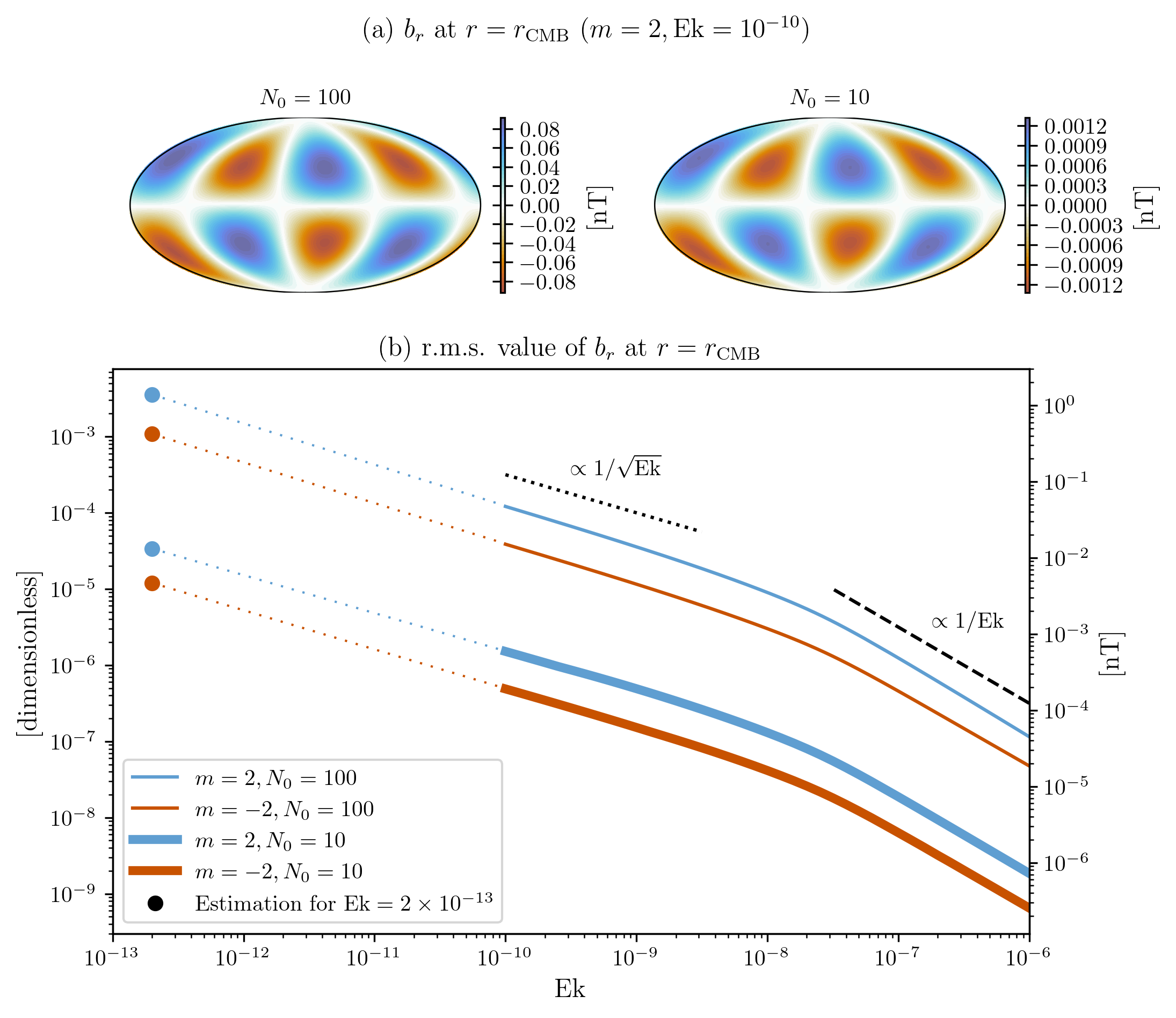}
\caption{Induced magnetic field by the core flow resulting from the radial components of libration given an outer stably stratified layer with $N_0 = 100$, $h = 0.1$, $r_\mathrm{C} = 0.7$, and $r_\mathrm{ICB} = 0.4$ and assuming a conductive layer at the base of the mantle with $c = 10^{-4}$ and $c' = 10^{-4}$. (a): R.m.s value of the radial magnetic field $b_r$ at the core-mantle boundary as a function of the Ekman number in response to the $m=2$ and $m=-2$ libration forcing. (b): Radial magnetic field at the core surface induced by the $m=2$ forcing component for $\mathrm{Ek} = 10^{-10}$.}
\label{fig:magnetic}
\end{figure}

Although our choice of background magnetic field is too weak to significantly alter the fluid motions (Appendix \ref{app:sensitivity}), the strong flow $\mathbf{u}$ generated by the radial libration forcing (Figure \ref{fig:flow_profile}(d)) in combination with the background field could induce a considerable magnetic field that harmonically oscillates with the libration frequency $\omega$, unlike the much weaker flow generated by the axial libration forcing. We compute this induced magnetic field $\mathbf{b}$ by first solving Equations \eqref{eq:momentum_nd}-\eqref{eq:incompressible_nd} without a background magnetic field and then using the resulting velocity field $\mathbf{u}$ as input to solve the induction equation \eqref{eq:induction_nd}, given a thin conducting layer at the bottom of the mantle with $c=10^{-4}$ and $c' = 10^{-4}$. The geometry of the induced magnetic field, computed for different values of the \BVfreq near the CMB, resembles the vector fields that generate it (Figure \ref{fig:magnetic}(a)). The North-South anti-symmetry is related to the dipolarity of the background magnetic field and the $Y_2^2$ shape is related to the flow structure near the boundary (Figure \ref{fig:flow_profile}(d)). The average r.m.s. value of the magnetic field increases with decreasing Ekman number (Figure \ref{fig:magnetic}(b)), which is expected since the flow velocity also increases with decreasing Ekman number. The induced magnetic field scales approximately as $1/Ek$ for $Ek > 10^{-8}$ and $1/\sqrt{Ek}$ for $Ek < 10^{-8}$. Assuming the validity of the scaling law to the low Ekman number expected in Mercury's core, we estimate the induced non-axisymmetric magnetic field strength at Mercury's core mantle boundary to be $\SI{1.4}{\nano\tesla}$ for $N_0 = 100$ and $m=2$ or about $\SI{0.5}{\percent}$ of the applied background magnetic field, neglecting non-linear effects. Alternatively, we can estimate this value using the magnetic Reynolds number $\mathrm{Rm} = \mathcal{U}\mathcal{L}/\eta$, where $\mathcal{U}$ and $\mathcal{L}$ again are typical velocity and length scales for the flow motion. Similar to the Reynolds number (Equation \eqref{eq:comp_reynolds}) we can compute this for our strong horizontal flow motion as:
\begin{equation}
\mathrm{Rm} = \frac{\max{(\mathbf{u}_\mathrm{tan})}\left(r_\mathrm{CMB} - \max_r{(\mathbf{u}_\mathrm{tan})}\right)}{\mathrm{Em}}~.\label{eq:magnetic_reynolds}
\end{equation}
using the magnetic Ekman number $\mathrm{Em}$ instead of $\mathrm{Ek}$. We can show that the time dependence of the magnetic field, $\delta_t{\mathbf{b}}$, is small over one oscillation period, so that the induction Equation \eqref{eq:induction_nd} reflects a balance between the induction term $\nabla \times (\mathbf{u} \times \mathbf{B}_0)$ and the dissipation term $\nabla^2 \mathbf{b}$. This balance implies an estimate for the induced magnetic field $\mathbf{b} = \mathrm{Rm}\mathbf{B}_0$, which using values of Mercury's outer core is equal to $\SI{0.7}{\nano\tesla}$ for $N_0 = 100$ and $m=2$, similar to the value we found above.

Recent reconstructions of Mercury's magnetic field, inferred from MESSENGER magnetic field observations, show that non-axisymmetric magnetic field components might contribute from $\SI{3}{\percent}$ \citep[][]{plattner2021mercury} up to $\SI{20}{\percent}$ \citep[][]{wardinski2021internal} to the r.m.s. signal of the radial magnetic field at the core surface. The estimated $Y_2^2$ components of the Mercury's stationary internal magnetic field (i.e. the $g_2^2$ and $h_2^2$ Gauss coefficients) are around $\SI{1}{\nano\tesla}$ \citep{wardinski2019correlated}, similar to the librationally induced periodic magnetic field we compute above for $N_0 = 100, \mathrm{Ek} = 2\times10^{-13}$ and $\alpha_2 = 10^{-4}$, although its value is uncertain due to MESSENGER's orbital geometry \citep[][]{anderson2012low,thebault2018time,toepfer2022reconstruction}. Knowledge on the internal magnetic field will be significantly improved by the BepiColombo mission according to \citet[][]{heyner2021bepicolombo}, that estimate an absolute difference of a few \SI{}{\pico\tesla} between reconstructed and actual Gauss coefficients up to $\ell = 6$ for the first year of this mission. If a robust reconstruction of the non-axisymmetric internal field BepiColombo data gives evidence of an 88-day periodic signal of the $g_2^2$ and $h_2^2$ coefficients, we propose that such a signal could be due to the librationally induced core flow, as other $\mathrm{Y}_2^2$ signals produced by the deeper core dynamo will evolve over much longer convective timescales. Such a detection might be possible since the period of the signal is four times shorter than BepiColombo's nominal life time of 1 Earth year \citep[][]{montagnon2021bepicolombo}. Moreover, the induced field at the spacecraft is not weakened due to attenuation in the stable layer, as is the case for fastly varying components of the dynamo field, since it is generated close to the CMB. If observed, the amplitude of the non-axisymmetric field components can then be used to constrain the strength of Mercury's top stratified layer, the Ekman number in the core and the triaxial shape of the core-mantle boundary. We nevertheless caution that the results obtained here are only valid in a laminar flow setting, even though, as we show above, the flow near the CMB is most likely turbulent, especially for strong stratification. An updated model that addresses those issues is required to see if the librationally induced flow is indeed capable of inducing a similar non-axisymmetric $\mathrm{Y}_2^2$ component in the magnetic field, as we suppose here. 

\section{Conclusions} \label{sec:conclusions}
We have studied the fluid motions in Mercury's outer core in response to the main 88-day longitudinal libration. We represent the libration of the mantle as the sum of one axial and two radial harmonic motions of the outer core boundary and numerically solve for the resulting flow using a fully 3D, linear spectral method. The viscous and electromagnetic torques acting on the boundary in our numerical model are much smaller than the total torque  driving the libration. The influence of the laminar outer core flow on the libration amplitude is thus very small and below the precision of the libration amplitude of current and future libration measurements and as such does not need to be considered in studies of Mercury's libration. 

We have also investigated how stable stratification near the outer boundary affects the core motions resulting from libration. Without stratification the librating motion of the boundary resonates with a nearby (in frequency) inertial mode so that the librationally induced core flow has the same spatial structure as that inertial mode. With a stratified outer part of the core, radial motions are strongly suppressed and resonances are no longer observed in our models for $N_0 > 0.01$. The radial flow is mostly converted into a horizontal flow near the core-mantle boundary, which becomes stronger as viscosity decreases. In the low viscosity regime expected in Mercury's fluid outer core, the tangential flow might become intense enough to induce an observable non-axisymmetric ($m=2$) structure in the magnetic field. 

If such an internal magnetic field structure with an $m=2$ character and an 88-day periodicity could be inferred from observations of Mercury's magnetic field, and recent papers \citep[e.g.][]{plattner2021mercury, wardinski2021internal} show that this might indeed be possible, then the strength of that magnetic field could inform on the \BVfreq near the core-mantle boundary as well as the Ekman number in the outer core and the triaxial shape of the core-mantle boundary. But as the induced magnetic field results are based on a model assuming laminar flow, and we expect the flow near the CMB to be turbulent, further study is needed to take non-linear interactions within the fluid into account and draw any definitive conclusions. 


\begin{acknowledgments}
The authors would like to thank two anonymous reviewers for their interesting and helpful comments that helped to improve the quality of the manuscript. The authors also wish to thank Véronique Dehant and Felix Gerick for many useful discussions early on in the writing of the manuscript. The research leading to these results has received funding from the European Research Council (ERC) under the European Union's Horizon 2020 research and innovation program (Synergy Grant agreement no. 855677 GRACEFUL), from the Belgian Federal Science Policy Office (BELSPO) for the BRAIN-be2.0 project STEM, and from the Belgian PRODEX program managed by the European Space Agency in collaboration with BELSPO for the project Planet Interior. 
\end{acknowledgments}

%
%
%
%
%


\appendix

\section{Poloidal-toroidal decomposition of the boundary conditions} \label{app:poltor_bound}
The boundary conditions on the velocity and magnetic field can also be projected on their respective poloidal and toroidal scalars. To establish no-slip conditions at the core-mantle boundary in terms of the poloidal $U$ and toroidal $V$ scalars we need to express the libration forcing $\mathbf{v}$ in those scalar functions. For the $m=0$ component given in Equation \eqref{eq:vel_phi}, this can be accomplished by writing the tangential velocity as $v_\phi = -\partial_\phi V$, so that:
\begin{equation}
U |_{r = r_\mathrm{CMB}} = 0~, 
\end{equation}
\begin{equation}
V |_{r = r_\mathrm{CMB}} = i  \omega r_\mathrm{CMB} \frac{\epsilon}{2} \mathrm{Y}_1^0 e^{i\omega t} + \mathrm{c.c.}~. \label{eq:no_slip0}
\end{equation}
Similarly, the radial velocity can be written as $v_r = \ell(\ell+1)r^{-1}U$ so that the $m=2$ forcing component, Equation \eqref{eq:vel_rad}, leads to:
\begin{equation}
U |_{r = r_\mathrm{CMB}} = \mp \omega {r_\mathrm{CMB}}^2 \frac{\alpha_2\epsilon}{\ell(\ell+1)} \mathrm{Y}_2^{\pm 2}e^{i\omega t} + \mathrm{c.c.}~, \label{eq:no_slip2_1}
\end{equation}
\begin{equation}
V |_{r = r_\mathrm{CMB}} =  0 .
\label{eq:no_slip2}
\end{equation}
At the inner core boundary the no-slip conditions imply $\mathbf{u} = 0$ (Equation \eqref{eq:no_slip}), so we set:
\begin{equation}
U = \partial_r U |_{r = r_\mathrm{ICB}} = 0~, \label{eq:noslip_pol} 
\end{equation}
\begin{equation}
V |_{r = r_\mathrm{ICB}} = 0~. \label{eq:noslip_tor}
\end{equation}

For the induced magnetic field at the core-mantle boundary we use the generalised thin-wall conditions as defined in \citet{roberts2010numerical}:
\begin{equation}
\left. (1 + c\ell) \partial_r F_{\ell m} +  \frac{\ell}{r_\mathrm{CMB}} F_{\ell m} + c'r_\mathrm{CMB}\left(1 + \frac{c\ell}{2}\right)\left[\partial_r^2 F_{\ell m} - \frac{\ell(\ell+1)}{{r_\mathrm{CMB}}^2}F_{\ell m}\right] \right|_{r = r_\mathrm{CMB}} = 0~, \label{eq:thinwall_pol}
\end{equation}
\begin{equation}
\left. G_{\ell m} + c' r_\mathrm{CMB} \partial_r G_{\ell m}  \right|_{r = r_\mathrm{CMB}} = 0~. \label{eq:thinwall_tor}
\end{equation}
The thin-wall conditions assume a thin electrically conducting layer between mantle and outer core with continuous magnetic boundary conditions on both interfaces. The exact conditions, Equation \eqref{eq:thinwall_pol} and \eqref{eq:thinwall_tor}, are recovered by taking the limit of the thin wall width $\delta$ to zero, while keeping the following "thin-wall parameters" $c$ and $c'$ constant (see \citet[][]{roberts2010numerical} or \citet[][]{guervilly2013effect} for details):
\begin{equation}
 c= \frac{\delta \mu_W}{r_\mathrm{CMB} \mu_F}~, \hspace{4cm}  c' = \frac{\delta \sigma_W}{r_\mathrm{CMB} \sigma_F} ~. \label{e1:cc1}
\end{equation}
Here $\mu_F$ and  $\sigma_F$ are the magnetic permeability and electrical conductivity of the outer core fluid and $\mu_W$ and $\sigma_W$ the permeability and conductivity of the conducting layer, that are unknown for Mercury. Unless the conductivities and permeabilities of the mantle and fluid are known, it is generally assumed that the magnetic permeabilities of both the conducting layer and the core fluid equal the vacuum permeability $\mu_0$ and that the conductivity in the bottom of the mantle is lower than the conductivity of the fluid, so that:
\begin{equation}
 c= \frac{\delta}{r_\mathrm{CMB}}~, \hspace{4cm}  c' = \frac{\delta \sigma_W}{r_\mathrm{CMB} \sigma_F} \leq 1 ~. \label{eq:cc2}
\end{equation}
As mentioned in the main text, the thin-wall condition can only be used when the width of the layer $\delta$ is much smaller than the generalised skin depth of the mantle, defined as $\mathcal{\delta}_\eta = \eta_W/(\omega+m\Omega)^{1/2}$ \citep[][]{guervilly2013effect}. In our model this demand is fulfilled when the control parameters $c$ and $c'$ satisfy:
\begin{equation}
cc' \ll \frac{\eta}{{r_\mathrm{CMB}}^2 (\omega + m\Omega)} ~. \label{eq:thinwall}
\end{equation}
where $\omega, m, \Omega^{-1}$ are the frequency, azimuthal wave number and rotation time scale governing the harmonic oscillations of the wall. Using the values in Table \ref{tab:parameters} we find:
\begin{equation}
cc' \ll 7.4 \times 10^{-7}~, \label{eq:thinwall2}
\end{equation}
and by Equation \eqref{eq:cc2} we have $c' \leq c$ so that:
\begin{equation}
c' \ll 2.7 \times 10^{-4}~. \label{eq:thinwall3}
\end{equation}
Realistically we don't assume that the width of the thin layer is much larger than $\SI{1}{\kilo\meter}$, in which case $cc' \leq 10^{-8}$ so the TWA should be valid for all models presented in this study. 

For ease of computation we assume that the inner core is insulating, so that the induced magnetic field can be written as a potential field $\mathbf{b} = \mathbf{\nabla} \phi$. The radial poloidal and toroidal scalar functions governing the induced magnetic field at the ICB then satisfy:
\begin{equation}
\left. \ell \frac{F_{\ell m}}{r_\mathrm{ICB}} - \partial_r F_{\ell m} \right|_{r = r_\mathrm{ICB}} = 0~, \label{eq:insulating_pol}
\end{equation}
\begin{equation}
G_{\ell m} |_{r = r_\mathrm{ICB}}  = 0~. \label{eq:insulating_tor}
\end{equation}

\section{Sensitivity of the results to different parameters} \label{app:sensitivity}
\begin{figure}
\centering
\plotone{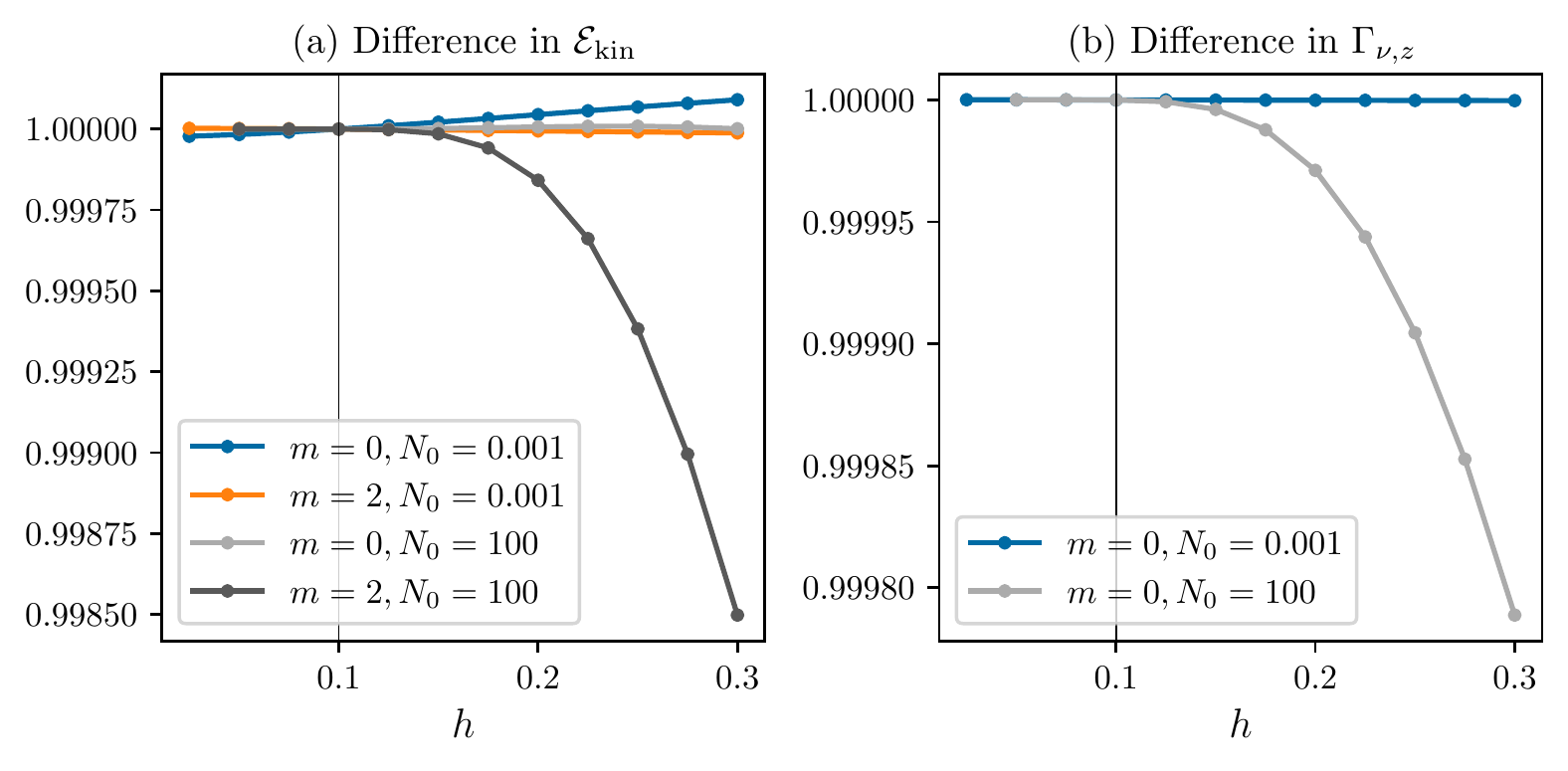}
\caption{Division of the total kinetic energy $\mathcal{E}_\mathrm{kin}$ and CMB viscous torque $\Gamma_{\nu, z}$ of the core flow, computed at different values of the smoothness parameter $h$, by a reference solution, denoted by the black vertical line.}
\label{fig:testh}
\end{figure}

\begin{figure}
\centering
\plotone{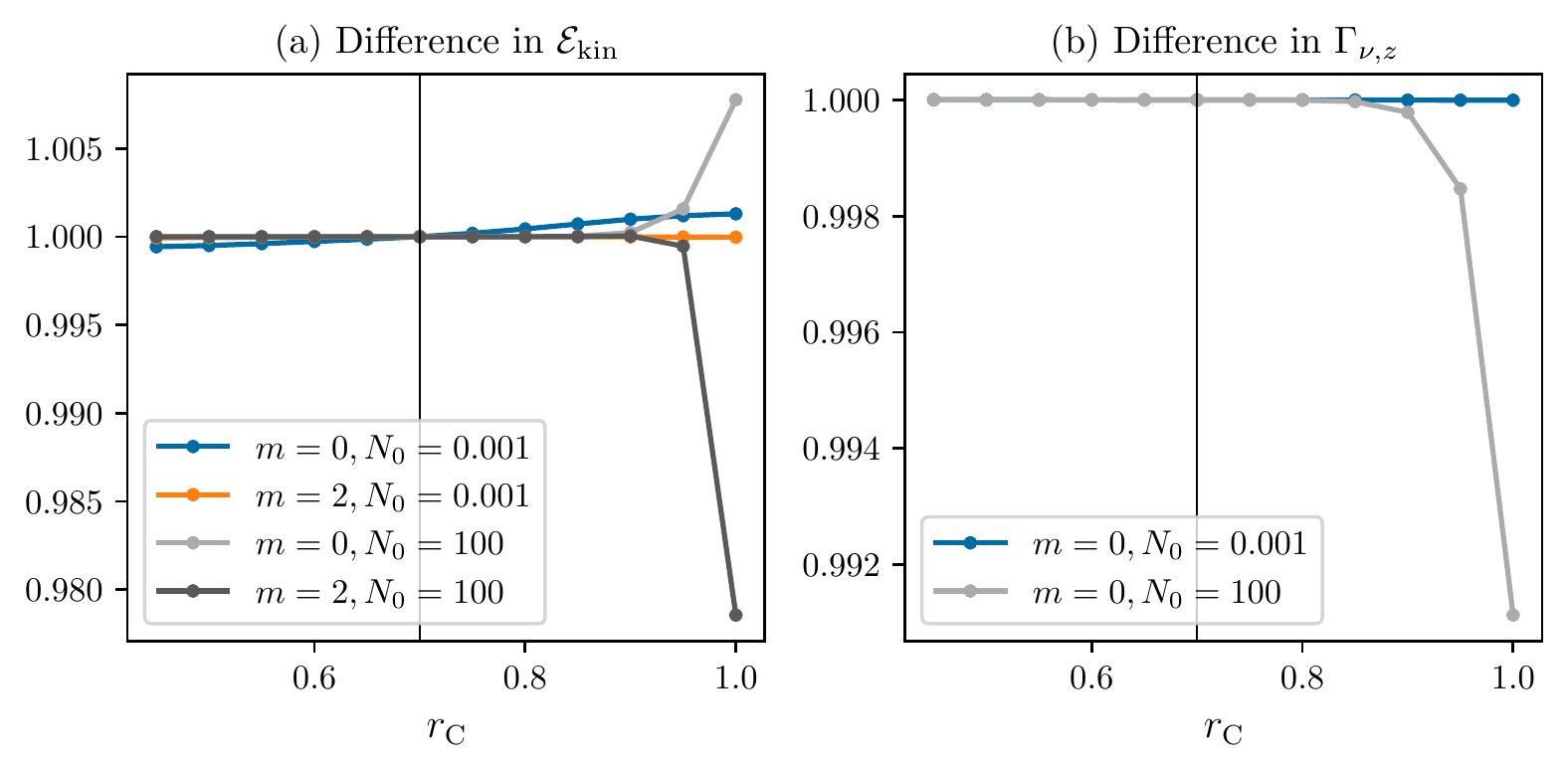}
\caption{Division of the total kinetic energy $\mathcal{E}_\mathrm{kin}$ and CMB viscous torque $\Gamma_{\nu, z}$ of the core flow, computed at different values of the convective core radius $r_\mathrm{C}$, by a reference solution, denoted by the black vertical line.}
\label{fig:testrc}
\end{figure}

\begin{figure}
\centering
\plotone{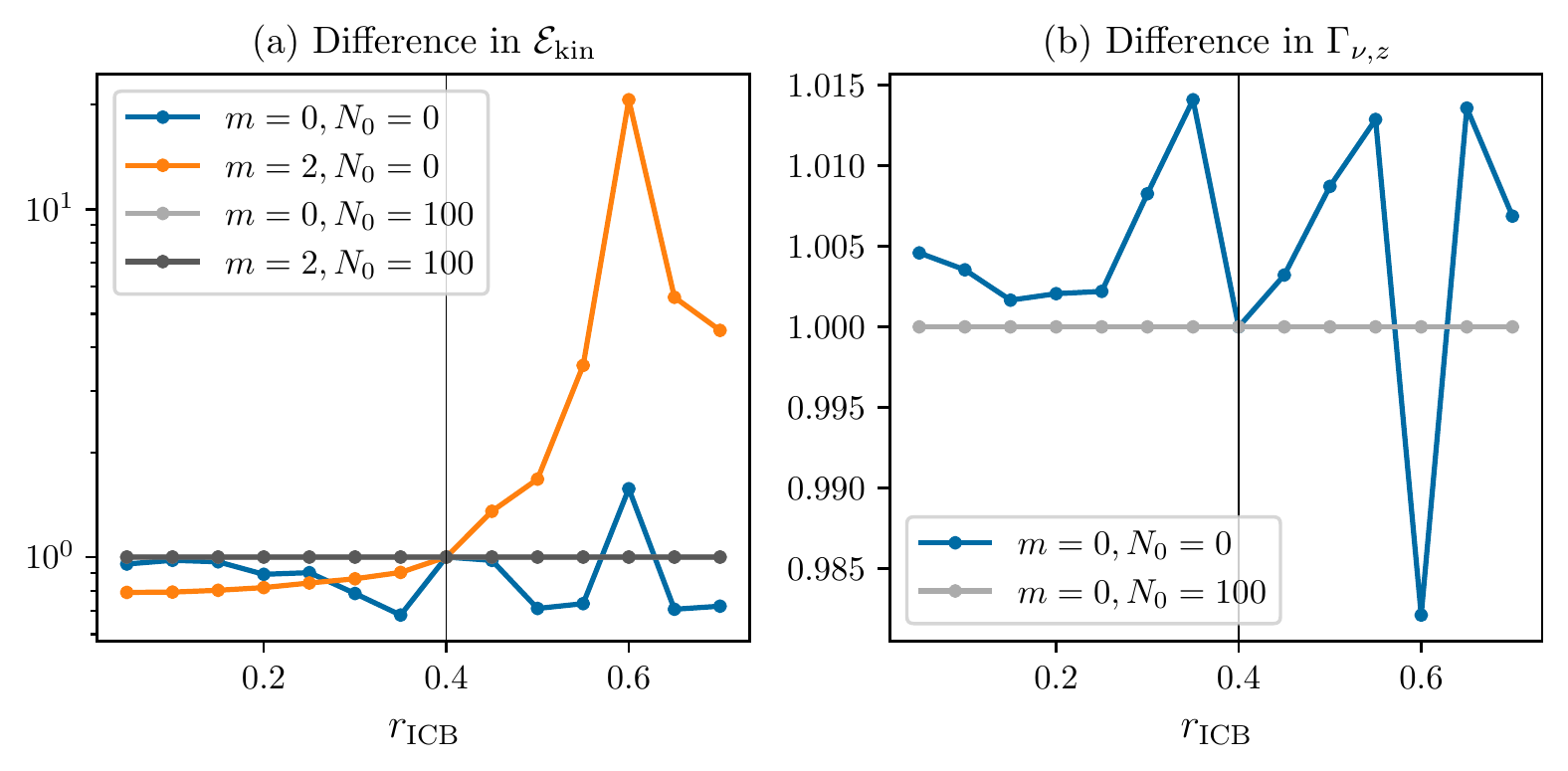}
\caption{Division of the total kinetic energy $\mathcal{E}_\mathrm{kin}$ and CMB viscous torque $\Gamma_{\nu, z}$ of the core flow, computed at different values of the inner core radius $r_\mathrm{ICB}$, by a reference solution, denoted by the black vertical line.}
\label{fig:testricb}
\end{figure}

\begin{figure}
\epsscale{1.15625}
\centering
\plotone{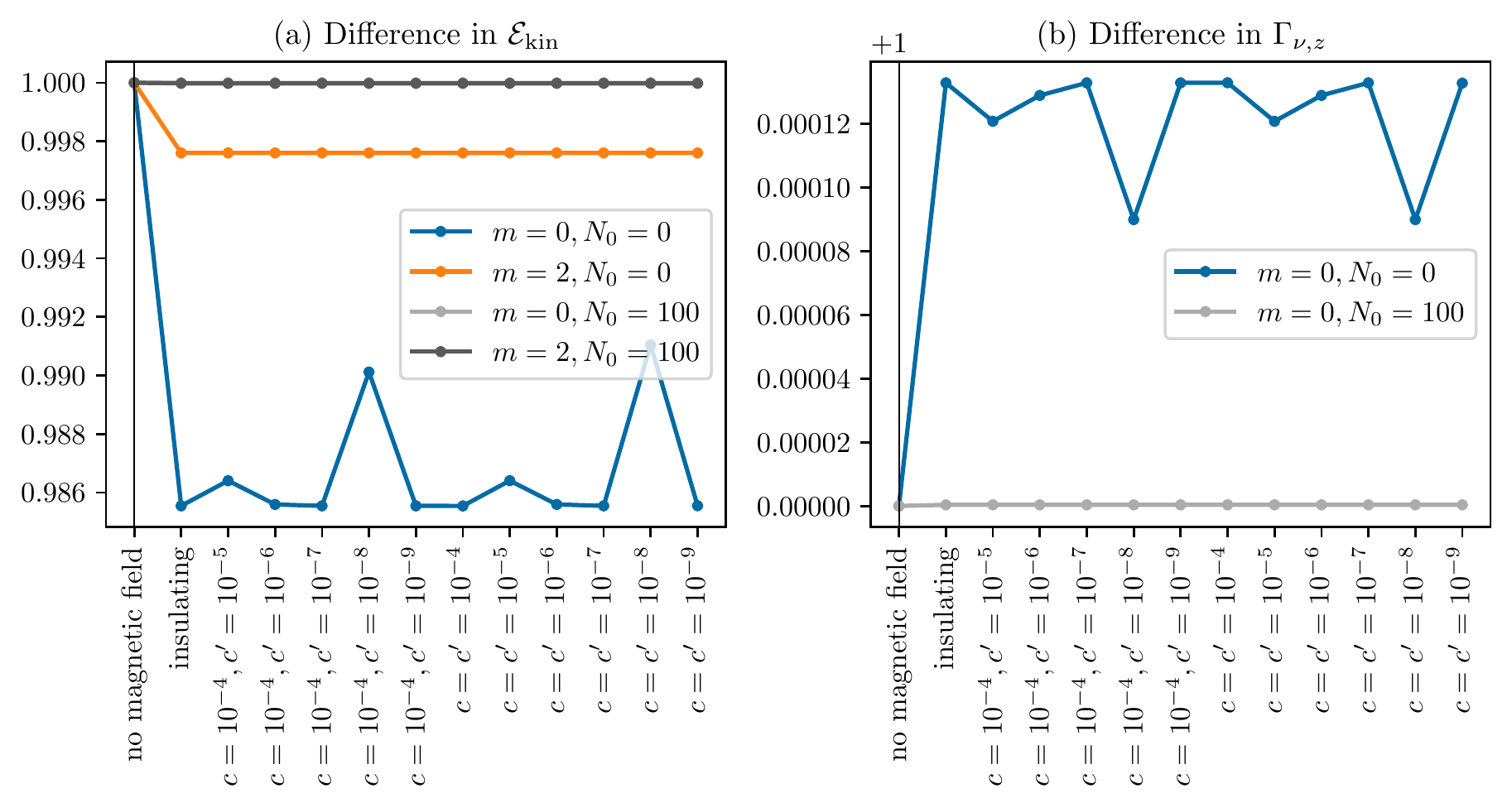}
\caption{Division of the total kinetic energy $\mathcal{E}_\mathrm{kin}$ and CMB viscous torque $\Gamma_{\nu, z}$ of the core flow, computed for different magnetic field conditions, by a reference solution, denoted by the black vertical line.}
\label{fig:testB}
\end{figure}

In this Appendix we show the sensitivity of our results in Sections \ref{sec:torque} and \ref{sec:results} to the smoothness parameter $h$, convective core radius $r_\mathrm{C}$, inner core radius $r_\mathrm{ICB}$, background magnetic field $\mathbf{B}_0$, and the magnetic boundary conditions on the inner core surface. For each parameter that we consider, we compute for different values of that parameter $\Gamma_{\nu, z}$, the axial viscous torque magnitude at the CMB, and $\mathcal{E}_\mathrm{kin}$, the total kinetic energy, while keeping all other parameter values fixed. The obtained torques and energies are respectively divided by the viscous torque and kinetic energy of a reference solution that is computed with the fixed parameter values $h=0.1$, $r_\mathrm{C}=0.7$, $r_\mathrm{ICB} = 0.4$ (as established in Section \ref{sec:method}) without a magnetic field ($\mathbf{B}_0 = 0$)(Figures \ref{fig:testh}-\ref{fig:testB}). In these figures the larger a test result deviates form one, the more that particular parameter value influences the results in Sections \ref{sec:torque} and \ref{sec:results}. In particular we examine four different cases, with little or no stratification ($N_0 = 0.001, 0$), with strong stratification ($N_0 = 100$), and for axial ($m=0$) and radial ($m=2$) libration forcing; which covers most of the cases that are considered in the aforementioned sections. 

We first note that our results are not overly sensitive to the smoothness parameter $h$, except when $h$ is large and the stratification strength $N_0$ is strong (Figure \ref{fig:testh}). In that case, however, the transition region becomes so wide that the \BVfreq near the core-mantle boundary doesn't yet approach its asymptotic value $N_0 = 100$ and is much lower. The observed differences in kinetic energy and viscous torque are then most likely caused by the difference in $N_0$ at the CMB, not the different shape of the \BV function in the bulk of the core. 
A similar thing happens when we vary the convective core radius $r_\mathrm{C}$ (Figure \ref{fig:testrc}); only for very high values of $r_\mathrm{C}$, when the \BVfreq near the CMB changes, do the strongly stratified solutions diverge from the reference solution. 
A different picture emerges when we change the inner core radius $r_\mathrm{ICB}$. When stratification is strong the value of $r_\mathrm{ICB}$ has no influence on either the kinetic energy or the viscous torque, but when the outer core is neutrally stratified especially the kinetic energy changes considerably for different values of the inner core radius (Figure \ref{fig:testricb}). We suspect that this happens because the inertial modes can be very sensitive to the inner core size \citep[see e.g.][]{tilgner2015rotational}. Accordingly for different inner radii the librationally induced core flow will resemble a different inertial mode with a different pattern of internal shear layers, thereby changing the total kinetic energy and the viscous torque acting on the CMB. 
Finally the kinetic energy and the viscous torque slightly change when a magnetic field is considered versus when there is no magnetic field, and for different parameters of the thin wall boundary condition but only without a stratified layer (Figure \ref{fig:testB}), again because in a neutrally stratified core, the inertial modes are more sensitive to different types of magnetic fields, although in this case the inertial modes depend very little on the background magnetic field and its boundary conditions, so that our results are only limitedly sensitive to the magnetic field parameters. If the stratification is strong the presence of a magnetic field doesn't change the results at all. 

In conclusion, solutions with a strong stably stratified outer layer are mostly insensitive to any of the variables considered here and, with the exception of the inner core radius, the same is true for solutions that assume a neutrally stratified core.

\section{Benchmarking the poloidal boundary condition}

\begin{figure}
\epsscale{1.15625}
\centering
\plotone{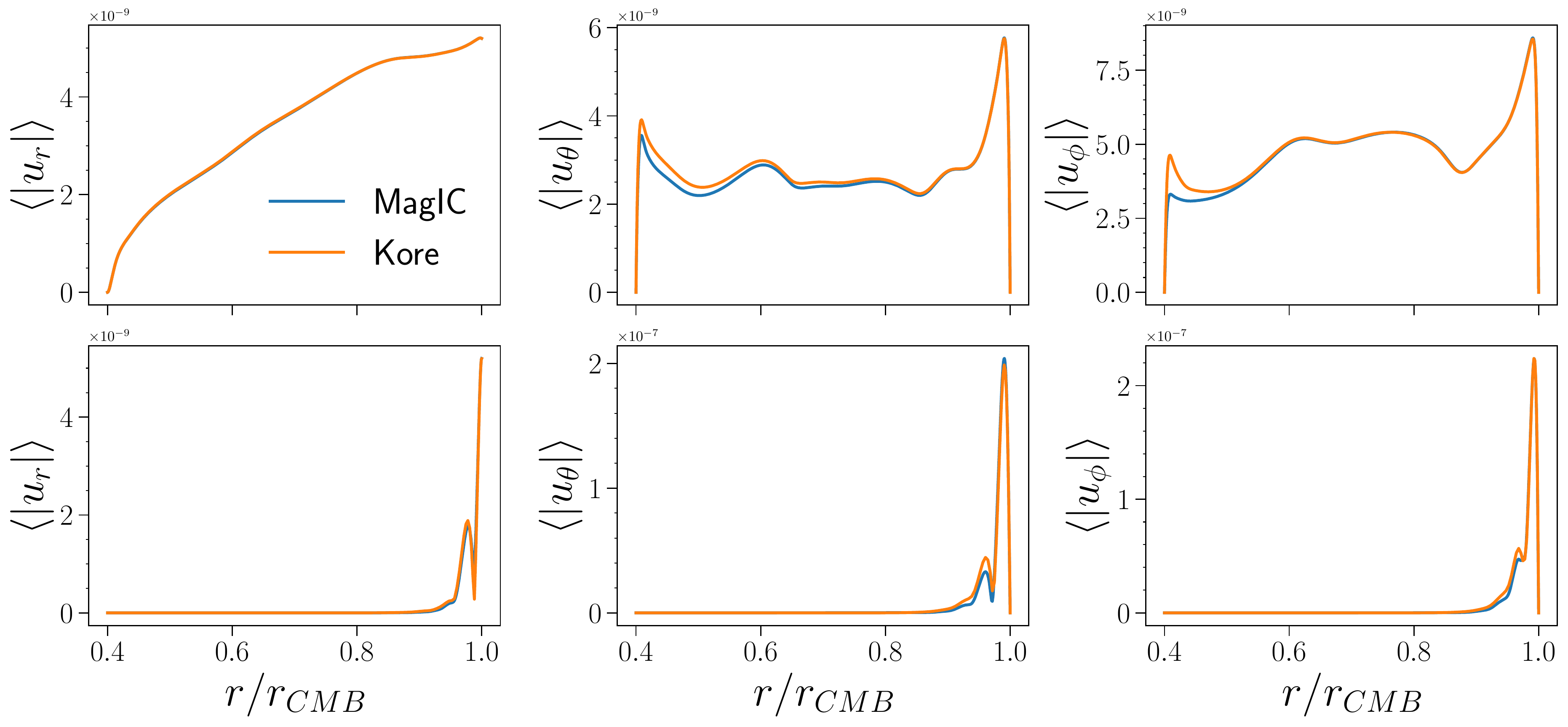}
\caption{Comparison of horizontally averaged profiles of radial ($\langle u_r\rangle$), latitudinal ($\langle u_\theta \rangle$) and azimuthal ($\langle u_\phi\rangle$) velocities from \texttt{MagIC} and \texttt{Kore}. The top panel shows comparison case 1 with no stable layer while the lower panel shows case 2 with a stably stratified layer.}
\label{fig:prof_comp}
\end{figure}

\begin{figure}
\epsscale{1.15625}
\centering
\plotone{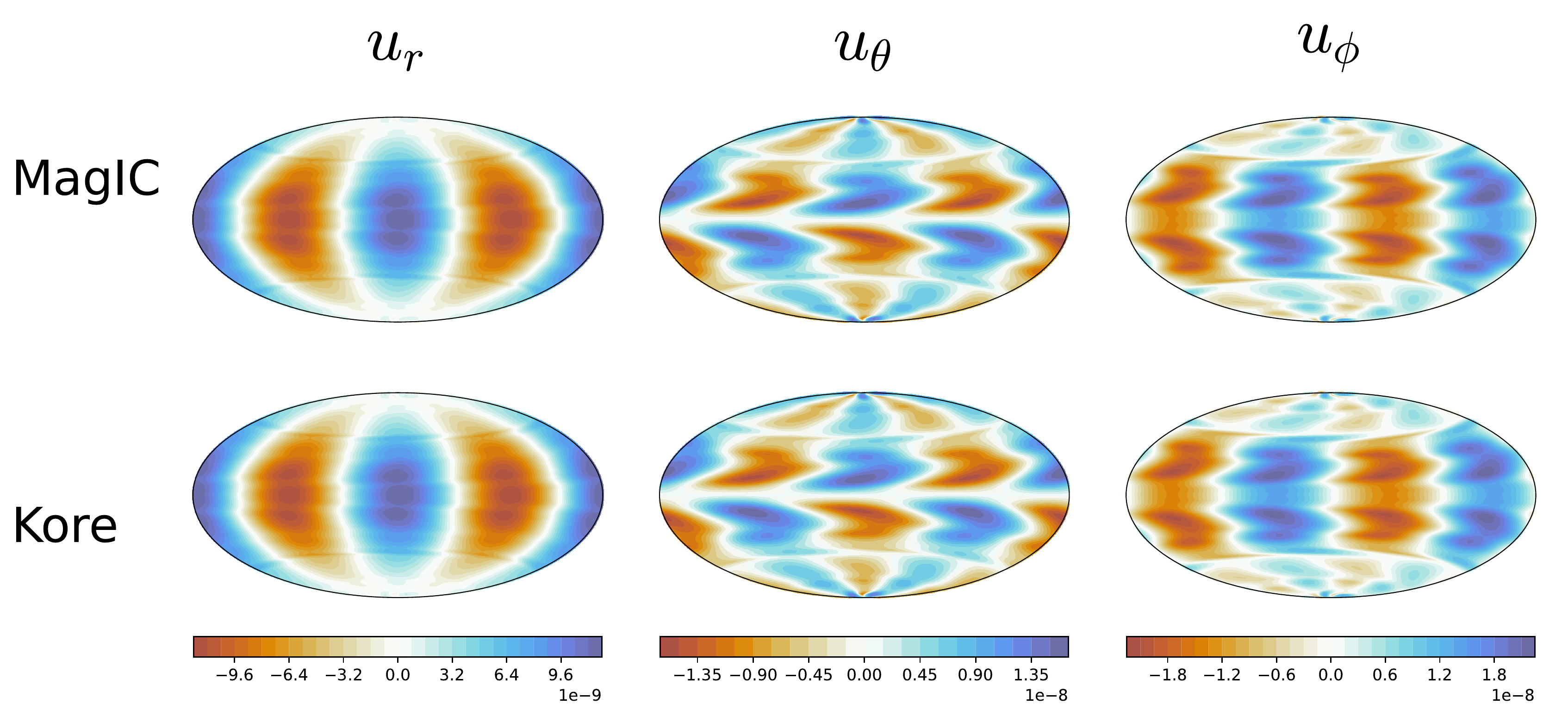}
\caption{Comparison of structure of modes at 99\% of outer boundary radius from the two codes \texttt{MagIC} and \texttt{Kore}. This figure shows comparison case 1 with no stable layer.}
\label{fig:surface_noSSL}
\end{figure}

\begin{figure}
\epsscale{1.15625}
\centering
\plotone{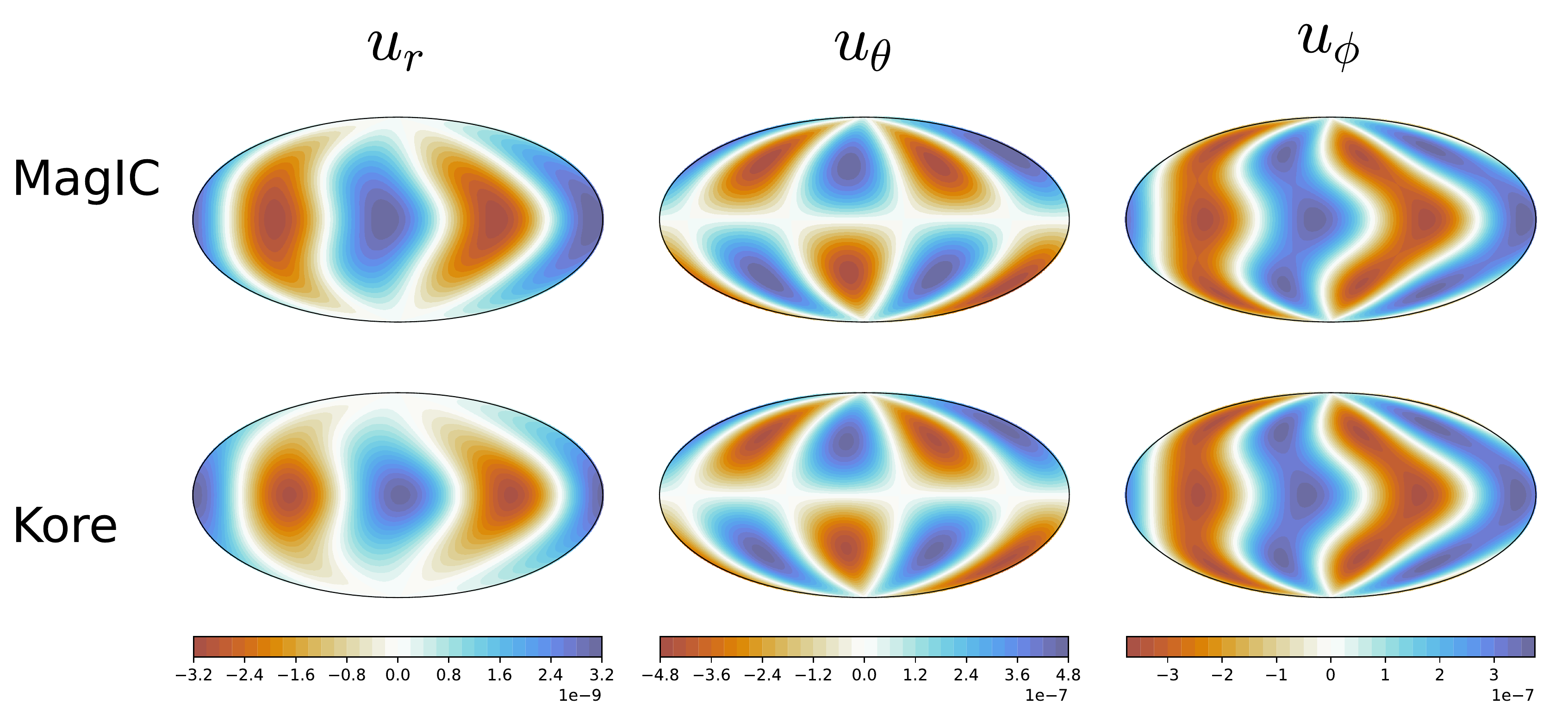}
\caption{Same as Figure \ref{fig:surface_noSSL} but for comparison case 2 with a stably stratified layer.}
\label{fig:surface_SSL}
\end{figure}

We benchmarked the poloidal boundary condition \eqref{eq:no_slip2_1} using the 3D MHD code \verb|MagIC| \citep{wicht2002inner, gastine2022magic} which uses the SHTns library \citep{schaeffer2013efficient,ishioka2018new}. \verb|MagIC| uses an expansion in spherical harmonics in the angular direction and for the radial derivatives, we used a fourth-order finite difference. MagIC was run in a linear mode restricted to only $m=2$ spherical harmonics. We tested two cases, a purely hydrodynamic simulation with an isothermal fluid and no stably stratified layer at $\mathrm{Ek}=3.6\times 10^{-6}$ (hereafter ``comparison case 1'') and a simulation using the temperature profile \eqref{eq:BV_profile} with with $N_0 = 100$ at $\mathrm{Ek}=3.6\times 10^{-5}$ (hereafter ``comparison case 2''). Figure \ref{fig:prof_comp} compares radial profiles of the horizontally averaged magnitudes of all three velocity components for the two cases, respectively, in its top and bottom panels. All the profiles match very closely with each other.
The structures of the modes are further compared in Figures \ref{fig:surface_noSSL} and \ref{fig:surface_SSL}. These figures show the structures of the three velocity components at 99\% of the outer boundary radius for the two comparison cases 1 and 2 without and with a stably stratified layer, respectively. The mode structures again show a remarkable similarity with each other.

\bibliography{bibliography}{}
\bibliographystyle{aasjournal}



\end{document}